\documentclass[useAMS,usenatbib]{mn2e}
\usepackage{amsmath}
\usepackage{amsfonts}
\usepackage{amssymb,graphicx}
\usepackage{pdflscape}
\usepackage{natbib}
\usepackage{journal_names}
\usepackage{longtable}
\usepackage{tabular}
\usepackage{supertabular,multicol}
\usepackage{subfigure}
\usepackage{booktabs}
\usepackage[titletoc]{appendix}
\usepackage{rotating}

\def\jnlref#1{{\rm#1}}
\def\aj{\jnlref{AJ}}
\def\araa{\jnlref{ARA\&A}}
\def\apj{\jnlref{ApJ}}
\def\apjl{\jnlref{ApJ}}
\def\apjs{\jnlref{ApJS}}
\def\ao{\jnlref{Appl.~Opt.}}
\def\apss{\jnlref{Ap\&SS}}
\def\aap{\jnlref{A\&A}}

\def\mnras{\jnlref{MNRAS}}
\def\nat{\jnlref{Nature}}

\def\pasj{\jnlref{PASJ}}

\def\procspie{\jnlref{Proc.~SPIE}}

\title[A deeper look at the X-ray point source population of NGC 4472]{A deeper look at the X-ray point source population of NGC 4472}
\author[T. D. Joseph, T. J. Maccarone, R. P. Kraft and G. R. Sivakoff] {T. D. Joseph$^{1,2,4}$\thanks{E-mail:
tana@ast.uct.ac.za; thomas.maccarone@ttu.edu; kraft@head.cfa.harvard.edu; sivakoff@ualberta.ca}, T. J. Maccarone$^{3,2}$\footnotemark[1], R. P. Kraft$^{4}$\footnotemark[1] and G. R. Sivakoff$^{5}$\footnotemark[1]\\
$^{1}$University of Cape Town, Rondebosch, 7701, Republic of South Africa\\
$^{2}$University of Southampton, Southampton, SO17 1BJ, United Kingdom\\
$^{3}$Texas Tech University, Lubbock, TX, USA \\
$^{4}$Harvard/Smithsonian Center for Astrophysics, 60 Garden Street, MS-4, Cambridge, MA 02138, USA\\
$^{5}$University of Alberta, Edmonton Alberta T6G 2E1, Canada}

\begin{document}

\date{Accepted ???. Received ???; in original form ???}

\pagerange{\pageref{firstpage}--\pageref{lastpage}} \pubyear{???}

\maketitle

\label{firstpage}

\begin{abstract}
In this paper we discuss the X-ray point source population of NGC\,4472, an elliptical galaxy in the Virgo cluster. We used recent deep \emph{Chandra} data combined with archival \emph{Chandra} data to obtain a 380\,ks exposure time. We find 238 X-ray point sources within 3.7$'$ of the galaxy centre, with a completeness flux, F$_{\rm X, 0.5-2\,keV}$= 6.3$\times10^{-16}$\,erg\,s$^{-1}$\,cm$^{-2}$. Most of these sources are expected to be low mass X-ray binaries. We finding that, using data from a single galaxy which is both complete and has a large number of objects ($\sim$\,100) below 10$^{38}$\,erg\,s$^{-1}$, the X-ray luminosity function is well fit with a single power law model. By cross matching our X-ray data with both space based and ground based optical data for NGC\,4472, we find that 80 of the 238 sources are in globular clusters. We compare the red and blue globular cluster subpopulations and find red clusters are nearly six times more likely to host an X-ray source than blue clusters. We show that there is evidence that these two subpopulations have significantly different X-ray luminosity distributions. Source catalogues for all X-ray point sources, as well as any corresponding optical data for globular cluster sources, are also presented here.
\end{abstract}

\begin{keywords}
X-rays: binaries -- X-rays: galaxies: individual: NGC\,4472 -- stars: low mass
\end{keywords}

\section{Introduction}

The \emph{Chandra X-ray Observatory}  \citep{2000SPIE.4012....2W} has made it possible to carry out detailed studies of low mass X-ray binaries (LMXBs) beyond the Local Group, allowing us to probe these systems in a wide variety of galactic environments. LMXBs can be formed either at the end point of stellar evolution. These binaries can also be produced in globular clusters (GCs) via N-body capture \citep{1975ApJ...199L.143C,1975MNRAS.172P..15F,1976MNRAS.175P...1H} or direct collisions between compact objects and red giant stars \citep{1987ApJ...312L..23V}. Yet, despite all we know about them, many unanswered questions about LMXBs still remain. Is the LMXB X-ray luminosity function universal across all galaxies? Is there a difference between LMXBs found in the field of the galaxy and those found in GCs? Why do these X-ray sources prefer red GCs over blue ones? The GC-LMXB relationship is of particular interest as these clusters appear to be a rich environment for LMXBs. It is necessary to study nearby elliptical galaxies and their GC systems to gain a better understanding of the issues related to LMXB formation and evolution.  

\par
NGC\,4472 is a massive \citep[8$\times10^{11}$ M$_{\odot}$;][]{2003ApJ...591..850C}, nearby elliptical galaxy that is falling into the Virgo cluster \citep[d = 16\,Mpc; e.g.][]{1999ApJ...521..155M}. It contains much less bright diffuse X-ray gas than other massive nearby ellipticals such as M\,87 and NGC\,1399, making it easier to probe fainter X-ray point sources within the galaxy. NGC\,4472 has a system of roughly 6000 globular clusters  \citep{2001AJ....121..210R} and is therefore an excellent environment in which to study the GC-LMXB connection.  

\par
 \citet{2002ApJ...574L...5K} studied the X-ray point source population of NGC\,4472 using a 40\,ks \emph{Chandra} observation and archival \emph{Hubble Space Telescope} (\emph{HST}) data  \citep[see also][]{2001AJ....121.2950K}. They considered only X-ray data from the \emph{Chandra} ACIS--S3 chip and excluded the central 8$''$ due to the high background in this region. A total of 144 X-ray sources were detected. This study found that the spatial distribution of the GCs, the GC-LMXBs and the field LMXBs were similar. They showed that 30 of the 72 LMXBs situated within the \emph{HST} fields were coincident with GCs. The subsequent work of \citet[][hereafter referred to as MKZ03]{2003ApJ...586..814M} found no statistically significant differences between the X-ray properties of the GC and field LMXBs. These findings could imply that a significant proportion of field LMXBs were created in GCs and then ejected from the cluster due to dynamical interactions or the destruction of the cluster.

\par
More recent deep \emph{Chandra} data has become available for NGC\,4472. By combining these data with archival observations we should be able to detect enough X-ray sources to carry out robust comparisons of various LMXB populations within the galaxy. 

\section{Observations and Data Analysis}

In this work we used both archival and recent Chandra observations of NGC\,4472 to study its X-ray point source population. All the data were taken with ACIS-S. The most recent data were taken in 2011 on February 14 (Obs ID 12889) and February 21 (Obs ID 12888); with exposure times of 140\,ks and 160\,ks respectively. Two shorter observations of 40\,ks each were also used, giving a total exposure time of 380\,ks. The more recent of the two  \citep[Obs ID 11274;][]{2010MNRAS.409L..84M} was taken on February 27 2010 and the oldest observation  \citep[Obs ID 321;][]{2002ApJ...574L...5K} was taken on June 12 2000. These data are summarised in Table\,\ref{NGC4472_obs}.

\begin{table*}
\centering
\begin{tabular}{|l|l|l|}
\toprule
Date &  Obs ID &  Exposure Time (ks) \\ 
\midrule
  2000-06-12 & 321 & 40 \\
  2010-02-27 & 11274 & 40 \\
  2011-02-14 & 12889 & 140 \\
  2011-02-21 & 12888 & 160 \\
\bottomrule
\end{tabular}
\caption[A list of \emph{Chandra} observations of NGC\,4472] {A list of all the \emph{Chandra} observations of NGC\,4472 used in this work.}
\label{NGC4472_obs}
\end{table*}

\par
All the data sets were first reprocessed with the {\sc ciao}\footnote[2]{http://cxc.harvard.edu/ciao4.4/threads/index.html} script \emph{chandra\_repro} using calibration data from March 2011. New level two event files were created for all the observations using this script. The data were then checked for background flares and these were eliminated. The event files and images were restricted to the well calibrated 0.5--5\,keV range using the {\sc ciao} command \emph{dmcopy}.

\par
Exposure maps were created for each CCD chip of each observation using the \emph{asphist}, \emph{mkinstmap} and\emph{mkexpmap} scripts. PSF files were produced using \emph{mkpsfmap} to determine the size of the PSF at each position on the images. The exposure maps and PSF file data were then used by the point source detection script \emph{wavdetect} with a false source detection probability threshold of $10^{-7}$ to correctly identify point sources in the images.

\par
All sources within 8$''$ of the galaxy centre were excluded to avoid the area of very high X-ray background, consistent with the work of  \citet{2002ApJ...574L...5K}. Furthermore, only sources within 3.7$'$ of the galaxy centre were included in this study, to avoid areas close to the CCD edges that were not adjacent to another CCD and to exclude the regions with substantially enlarged PSF. Thus we only include sources from ACIS-S3 and ACIS-S4 in this this work. 


\par
We obtained a counts-to-energy conversion factor by comparing the fluxes of two bright sources with their respective net count rate. The spectrum of a bright source that had a disk blackbody (diskbb) continuum with an inner disk temperature, $k$T = 1.5\,keV, and one that had a power law (PL) continuum with $\Gamma$=1.7 were analysed in ISIS and fluxes and flux errors were obtained. The respective net count rates (net counts per source divided by the net exposure) and the corresponding errors were obtained from \emph{wavdetect}. The counts-to-energy conversion factor was then calculated by dividing the flux obtained through spectral fitting by the net count rate of each source. These conversion factors were then used to calculate the X-ray fluxes for both diskbb and PL models for the rest of the X-ray point sources using their respective count rates. The errors for the flux values were calculated by using the flux errors obtained from ISIS for the two bright sources mentioned above and the errors on the count rate obtained from \emph{wavdetect} and propagating them in the standard way.

\par
A 4\,$\sigma$ completeness limit was calculated for the combined 380\,ks data set. Using the smoothed background image of the data set created in \emph{wavdetect}, the number of background counts at each source position was determined using the a region enclosing 90\,\% of the the source energy. This gives the background sensitivity as a function of position. The 4\,$\sigma$ upper limit was then calculated and these background counts were then converted to flux values using the counts-to-energy conversion factor. It was found that the diffuse X-ray emission in the central region of the galaxy dominated the background sensitivity and the completeness limit was determined based on the flux values in that region.  

\par
We made use of the {\sc topcat}  programme \citep{2005ASPC..347...29T} to search for sources coincident with GCs found in both ground  \citep{2001AJ....121..210R} and space based \citep{2009ApJS..180...54J} optical observations. \citet[][hereafter referred to as RZ01]{2001AJ....121..210R} surveyed the GC system of NGC\,4472 using \emph{B, V} and \emph{R} filters. The survey field covers the central 0.66$'$--23$'$ region of the galaxy and 1465 GCs were detected. 


\par
\citet{2009ApJS..180...54J} surveyed the GC system of the inner region of NGC\,4472 using the F475W and F850LP bandpasses as part of their \emph{Virgo Cluster Survey} (VCS). These bandpasses are approximately equivalent to the Sloan \emph{g} and \emph{z} filters respectively and will be referred to as such in the rest of the is work. The VCS covers the central 3.3$' \times$\,3.3$'$ of the galaxy  \citep[see ][]{2004ApJS..153..223C}; 764 GCs were detected in this region of NGC\,4472. 

\par
MKZ03, using Obs ID 321, found 30 X-ray sources coincident with GCs. They used archival \emph{HST} observations that covered central region as well as the halo of NGC\,4472 \citep[see Fig.\,1 in ][]{2002ApJ...574L...5K} and used $V$ and $I$ filters \citep[see][]{2001AJ....121.2950K}. We cross matched the GC source list of MKZ03 with our source catalogue using the \emph{Chandra} coordinates to try to identify more GC X-ray sources.

\par
We estimated the boresight corrections for the X-ray positions by calculating the weighted mean offset between obvious initial matches of the optical and X-ray data. The RZ01 data had offsets of -0.25$''$ (RA) and -0.50$''$ (DEC), while the VCS positions had offsets of 0.18$''$ and 0.10$''$ for RA and DEC respectively. These mean offsets were then applied to the X-ray positions.  

\par
We experimented with several different matching radii for the RZ01 and VCS catalogues. For the VCS catalogue, it was found that a matching radius of 0.5$''$ produced only 36 matches between X-ray and optical sources whereas a 0.75$''$ matching radius produced nearly twice as many. A 1$"$ matching radius yielded as many matches as the 0.75$"$ matching radius and thus we adopted the latter in our work.

\par
For the RZ01 catalogue, we tried various matching radii in the range 0.75$''$--2$''$. We found that a matching radius of 1.25$''$ provided a good number of matches without being so large as to include non-coincident sources.

\par
The false match probability for matches between the optical and X-ray data were determined by shifting each X-ray source position by a distance much greater than the matching radius and much smaller than the field of view and then seeing how many matches were found. In this case, the shift value used was 10$''$.

\section{Results and Discussion}

\subsection{X-ray point source population}

\begin{figure*}
\centering
\includegraphics[width=0.5\textwidth]{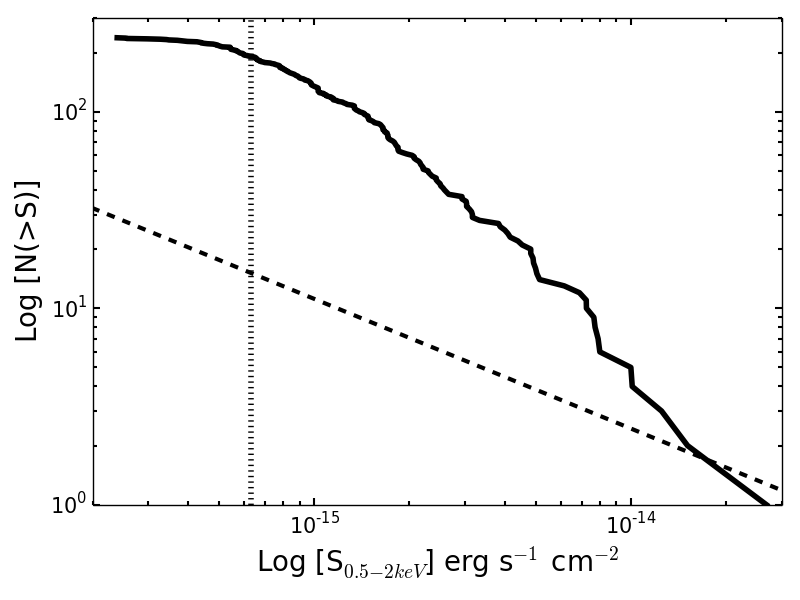}
\caption[The log\,N--log\,S plot for the X-ray point source population of NGC\,4472]{The log\,N--log\,S plot for the entire X-ray point source sample. The dotted vertical line indicates the 0.5--2\,keV completeness limit and the dashed line indicates the number of background sources as a function of flux \citep[see Eqn.\,1 in ][]{2001ApJ...562...42T}.}
\label{LogN_LogS}
\end{figure*}

We find 238 sources within 3.7$'$ of the galactic centre, 70 of which are bright enough to be detected in all four observations. 191 sources are found to be brighter than 6.3$\times10^{-16}$\,erg\,s$^{-1}$\,cm$^{-2}$ (0.5--2\,keV; L$_{\rm X,0.5-5\,keV}$=3.7$\times 10^{37}$\,erg\,s$^{-1}$), the 4\,$\sigma$ completeness limit for the combined data set. We expect roughly 10 of the X-ray sources to be background AGN at this completeness limit  \citep{2001ApJ...562...42T}. Data for AGN counts are generally given for a soft band (0.5--2\,keV) and a harder band. Since it was not possible to find AGN XLF data present specifically for the 0.5--5\,keV band we use in the rest of the paper, we present the 0.5--2\,keV data for for the AGN data as well as our own X-ray point source data for NGC\,4472. We also note that the soft band AGN slope is shallower than the hard band slope estimates, thus using the 0.5--2\,keV AGN data provides a slight overestimate of the number of expected background AGN. These data are summarised in Fig.\,\ref{LogN_LogS}. 

\par
A list of the X-ray luminosities of the sources can be found in Table\,\ref{source_list}. A sample of this table is shown in Table\,\ref{sample_table}. 

\begin{table*}
\tiny
 \begin{tabular}{|l|l|l|l|l|l|l|l|l|l|l|}
 \toprule

RA &
DEC &
Flux$_{\rm PL}$ &
Flux$_{\rm PL}$ err &
Flux$_{\rm PL}$ &
Flux$_{\rm PL}$ err &
Flux$_{\rm DBB}$ &
Flux$_{\rm DBB}$ err &
Flux$_{\rm DBB}$ &
Flux$_{\rm DBB}$ err &
Flag \\
\midrule
degrees &
degrees &
0.5--5\,keV &
0.5--5\,keV &
0.5--2\,keV &
0.5--2\,keV &
0.5--5\,keV &
0.5--5\,keV &
0.5--2\,keV &
0.5-2\,keV & 
 \\
\midrule
  187.49977 & 8.02063 & 1.456E-14 & 1.724E-15 & 7.641E-15 & 9.052E-16 & 1.536E-14 & 1.829E-15 & 6.248E-15 & 7.436E-16 & \\
  187.49804 & 8.02989 & 9.137E-16 & 2.96E-16 & 4.795E-16 & 1.553E-16 & 9.639E-16 & 3.125E-16 & 3.921E-16 & 1.271E-16 & \\
  187.49757 & 7.99843 & 1.097E-15 & 2.586E-16 & 5.759E-16 & 1.357E-16 & 1.158E-15 & 2.732E-16 & 4.709E-16 & 1.111E-16 & \\
  187.49175 & 7.96768 & 2.955E-15 & 7.334E-16 & 1.551E-15 & 3.849E-16 & 3.118E-15 & 7.746E-16 & 1.268E-15 & 3.151E-16 & \\
  187.4915 & 7.96826 & 1.831E-15 & 3.881E-16 & 9.61E-16 & 2.037E-16 & 1.932E-15 & 4.101E-16 & 7.858E-16 & 1.668E-16 & \\
  187.49098 & 7.99121 & 1.57E-15 & 3.028E-16 & 8.242E-16 & 1.589E-16 & 1.657E-15 & 3.201E-16 & 6.739E-16 & 1.302E-16 & \\
  187.49021 & 7.9924 & 1.708E-15 & 3.068E-16 & 8.962E-16 & 1.611E-16 & 1.801E-15 & 3.245E-16 & 7.328E-16 & 1.32E-16 & GC\\
  \bottomrule
 \end{tabular}
\caption[A sample of the table of the X-ray fluxes of X-ray sources in NGC\,4472]{A sample of the table of X-ray fluxes of the sources within 3.7$'$ of the galaxy centre. The fluxes are in units of erg\,s$^{-1}$\,cm $^{-2}$. The PL and DBB subscripts denote the spectral model conversion factor used to calculate the flux. The "err" designation denotes flux errors. The flag column identifies sources that are interesting and/ or for which caution must be used when analysing the source properties. The key is as follows: bkg = source in an area of high background, c = \emph{wavdetect} has misidentified or not detected the source in one of the observations, cg = the source is in or very close to a chip gap in one of the observations and GC = the source is associated with a globular cluster.}
\label{sample_table}
\end{table*}

\par
We have constructed the X-ray luminosity function (XLF) for NGC\,4472 using multi-epoch data. One might therefore believe that the population of transient X-ray sources in the galaxy would lead to an incorrect XLF. However, \citet{2003ChJAS...3..257G} point out that XLFs created from \emph{Chandra} observations of the same galaxy taken at different times and even from different parts of the galaxy were not significantly different from each other. Thus we expect that combining observations of NGC\,4472 taken at different time, but covering the same area to create one XLF would not lead to an incorrect XLF. \citet{2011ApJ...735...26S} also found that source variability did not change the slope or normalisation of an XLF created from merged data sets taken at different epochs compared to XLFs created from individual observations.

\par
We fit a power law function to the differential XLF, using only the completeness limited luminosity data. The function has the form $\frac{dN}{dL_{\rm X}}=kL_{\rm X}^{-\beta}$ \citep[see e.g.][]{2004MNRAS.349..146G}. To account for the background AGN contamination component we calculate the expected number of background AGN using the luminosity function from \cite{2001ApJ...562...42T} and subtract this contribution from the observed XLF. We used the {\sc xspec} spectral fitting tool \citep{1996ASPC..101...17A} to fit the power law model to the XLF data, using the Cash statistic.

\par
\begin{figure*}
\centering
\includegraphics[width=0.5\textwidth]{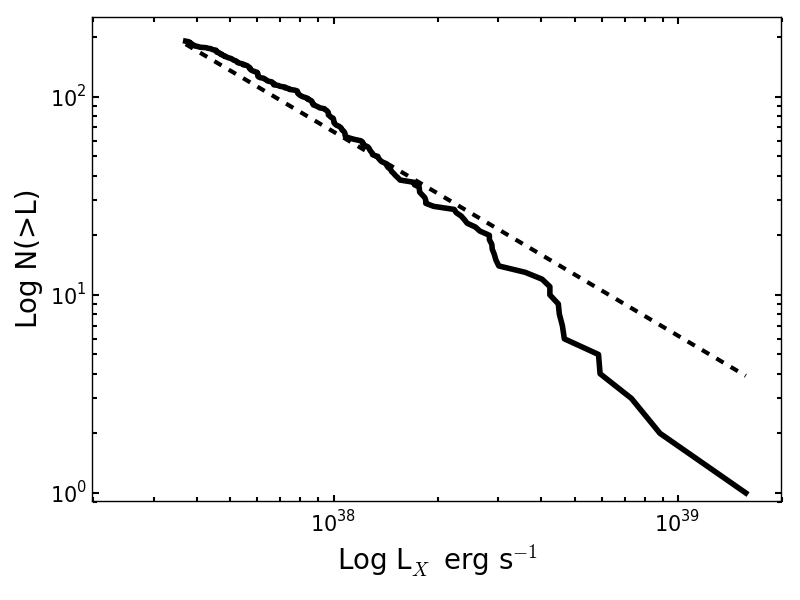}
\caption[The X-ray luminosity function for the X-ray point source population of NGC\,4472 with the best fit function over-plotted]{The X-ray luminosity function for the completeness limited X-ray point source sample (sold line) with the best fit function over plotted (dotted line). The luminosity is for the 0.5--5\,keV range.}
\label{XLF_slope}
\end{figure*}

We find that the best fit parameters for the slope and normalisation are $\beta=2.03^{+0.16}_{-0.15}$ and $k=68.53^{+8.95}_{-8.28}$ respectively. This slope is in good agreement with the results of other studies of the XLF for individual ellipticals as well as the average XLF for an ensemble of elliptical galaxies, where $\beta$ is typically found to be $\sim2$ \citep[e.g.][]{2004MNRAS.349..146G, 2004ApJ...611..846K,2006ApJ...652.1090K}. The cumulative XLF was constructed by integrating the differential XLF and is shown in Fig.\,\ref{XLF_slope}.

\par
The {\sc xspec} \emph{goodness} command was used to calculate the goodness-of-fit, using the Anderson-Darling test statistic, as this test is sensitive to differences between the model and the data at the edges of the  distribution \citep[see e.g.][]{2016MNRAS.458.3633M}. We ran 10\,000 simulations with the \emph{goodness} command to determine the null hypothesis probability. We find that our XLF data is well fit with a single power law model, with a null hypothesis probability of 0.24.

\par
Other studies have found that the XLF for LMXBs in elliptical galaxies are better fit with a broken power law function due to a flattening of the luminosity distribution at low luminosities \citep[L$_{\rm X}\lesssim 4\times 10^{37}$\,erg\,s$^{-1}$; e.g.][]{2006A&A...447...71V}. \citet{2004MNRAS.349..146G} suggested that this flattening is a ubiquitous property of the XLF for LMXBs. In the case of NGC\,4472 itself, \citet{2002ApJ...574L...5K} found that the XLF required a broken power law fit, with a break at L$_{\rm X}= 3\times 10^{38}$\,erg\,s$^{-1}$. This break is also seen in the XLF of other galaxies and it was suggested that this break corresponds to the transition in the XLF from neutron star to black hole binary luminosities \citep{2000ApJ...544L.101S}.

\par
We also fit the XLF with a broken power law model, using the same fit statistic and AGN contamination component as for the single power law. The best fit parameters for the broke power law are as follows: the slope values are 1.34$^{+0.43}_{-0.55}$ and 2.60$^{+0.55}_{-0.37}$ for the low and high luminosity range of the XLF, with a normalisation of 93.99$^{+34.30}_{-19.43}$. The break energy is is found to be 1.19$^{+0.83}_{-0.36}$\,keV. The \emph{goodness} command was used again with the Anderson-Darling test statistic; the null hypothesis was was to be 0.04\%. This result indicates that the broken power law model is overfitting the data and the added complication of a broken power law model is not justified.

\par
Therefore, since we find an acceptable fit to the data with only a single power law model, we do not consider a broken power law to be necessary to describe the XLF of NGC 4472. Indeed, \citet{2006ApJ...652.1090K} showed that the flattening of the XLF slope at low luminosities is not seen in their ensemble of galaxy XLFs. They caution against the acceptance of the flattening as a general characteristic of the XLF across all galaxies, since factors such as incompleteness and multiple X-ray point source populations may not have been taken into account in other studies. We note that NGC 4472 does not have multiple point source populations and we only use completeness limited sample, with $\sim$\,100 sources less luminous than 10$^{38}$\,erg\,s$^{-1}$, to construct our XLF. 

\subsection{Globular cluster LMXB population}
We find that a total of 80 of the 238 detected X-ray sources are associated with GCs, 68 with fluxes above the X-ray completeness limit. The X-ray and optical parameter values of these 80 sources can be found in Table\,\ref{GC_source_list}. A total of 233 X-ray sources lie within the RZ01 survey field and of these 40 were found to have a match with a RZ01 GC, with a subsample of 35 with fluxes above the X-ray completeness limit. The VCS field included 210 X-ray sources and 56 matches were found with this data, 46 of which have fluxes above the X-ray completeness limit. Cross matching the MKZ03 GC-LMXB catalogue yielded 24 sources that were matched to a X-ray source in our catalogue, with 23 sources above the X-ray completeness limit. Three of these MKZ03 GC matches were not found in either the VCS or RZ01 catalogues. We estimate that there would be two and five false matches in the RZ01 and VCS samples respectively. 

\par
For the purposes of our analysis, we apply X-ray and optical (see \S\,3.2.2) completeness cuts to our GC sample. A comparison of the RZ01 and VCS catalogues show that only 36\,\% of RZ01 GC sources are also found in the VCS catalogue in their overlapping region. Thus we also limit our analysis to field and GC sources that fall within the area covered by the VCS catalogue to ensure a cleaner sample of both GC and field sources for comparison. After applying these three cuts, we have a sample of 44 GC and 89 field LMXBs.

\subsubsection{Field versus globular cluster LMXB populations}

We compare the GC and field LMXB populations to ascertain whether the the two populations have significantly different X-ray properties. We find that there is not statistically significant difference between the mean 0.5--5\,keV luminosity for the GC-LMXB 1.1$\pm21.1\times 10^{38}$\,erg\,s$^{-1}$ and field LMXB populations 1.2$\pm1.3 \times 10^{38}$\,erg\,s$^{-1}$. These values and findings are consistent with those found by \citet{2002ApJ...574L...5K} for their smaller sample of NGC\,4472 LMXBs. 

\par
\begin{figure*}
\centering
\includegraphics[width=0.5\textwidth]{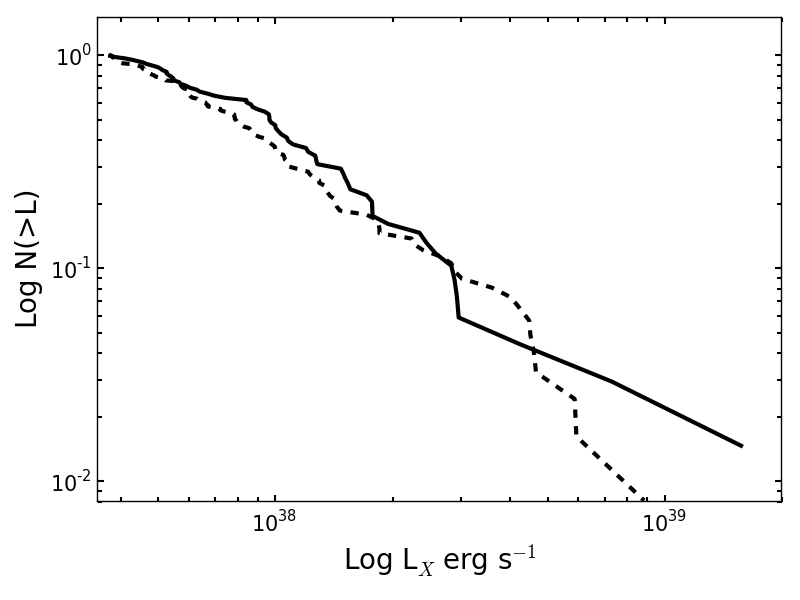}
\caption[The normalised log\,N--log\,L plots for the GC and field populations of NGC\,4472]{The log\,N--log\,L plots for the GC (solid line) and field (dashed line) populations. The number of sources in each sample has been normalised to one for ease of comparison.}
 \label{GC_vs_field}
\end{figure*}

The Anderson-Darling (AD) test was performed on the luminosities of the GC and field samples to quantify any differences between the two distributions. We find that the probability that the GC and field LMXB luminosities are drawn from the same sample is at least 25\,\%. The findings of previous studies of other elliptical galaxies \citep[see e.g.][]{2006ApJ...647..276K} as well as that of the previous study of NGC\,4472 \citep{2002ApJ...574L...5K} do not find any significant difference between the two luminosity distributions. Fig.\,\ref{GC_vs_field} shows the normalised log\,N--log\,S plots for both the GC and field LMXB populations.

\par
However, we note that our completeness limit is 3.7$\times 10^{37}$\,erg\,s$^{-1}$ and therefore we do not sample the same low luminosity range as other studies. For example, using samples with completeness limits $\lesssim 10^37$\,erg\,s$^{-1}$, \citet{2009ApJ...701..471V}, \citet{2009ApJ...703..829K} and \citet{2011A&A...533A..33Z} find that the luminosity functions for field and GC-LMXB populations differ significantly for sources with L$_{\rm X} \lesssim 4\times 10^{37}$\,erg\,s$^{-1}$, with fewer faint sources in the GC population. It is then perhaps not surprising that we do not see the flattening of the XLF slope at low luminosities since our completeness limited sample contains very few sources fainter than 4$\times 10^{37}$\,erg\,s$^{-1}$.

\par
In addition, \citet{2014ApJ...789...52L} and \citet{2016ApJ...818...33P} find a flatter slope for the XLF of the GC-LMXB population compared to that of the field population, which implies a difference between the XLFs at the bright end of the distribution as well. We performed an AD test on the just the bright (L$_{\rm X}> 10^{38}$\,erg\,s$^{-1}$) end of our GC and field luminosity distributions and find that the the probability that the GC and field bright end XLFs are statistically similar is at least 25\%. Thus our findings are that the field and GC-LMXB luminosity distributions are statistically indistinguishable, which is in contrast to the findings of other studies.

\par
The radial distance distributions of the GC and field LMXB populations were also analysed. For the GC source list, both X-ray and optical completeness was taken into account (see next section for optical completeness data). The average radial distance of field and GC sources was found to be 1.3$' \pm 0.7'$ and 1.5$' \pm 0.6'$ respectively. We perform an AD test and find the probability that the two radial distance distributions are drawn from the same sample to be 14\,\%.This result is similar to that of \citet{2002ApJ...574L...5K} who found that the probability that the GC and field sources were drawn from the same radial distribution population could not be ruled out at greater than the 30\,\% level.

\subsubsection{Red versus blue globular cluster subpopulations}

We compare the red and blue GC subpopulations to try and determine the role that GC colour, and hence metallicity, plays in LMXB formation. For some of the analysis, we combine our VCS, RZ01 and MKZ03 GC samples. This does not present a problem, even though not all GC sources have RZ01 or MKZ03 counterparts and vice versa. Colour bimodality is an intrinsic property of GC systems \citep{1998CAS....30.....A}. Therefore, for analysis that only requires the source to be labelled blue or red, it does not matter which photometric system was used to measure the colour of the GC. Once a GC colour has been determined and the GC has been assigned the red or blue qualitative property, we can use that to carry out further analysis. The colour bimodality of the GC system of NGC\,4472 is well known and has been quantified by VCS, RZ01 and MKZ03. 

\par
\begin{figure*}
\centering
\includegraphics[width=0.5\textwidth]{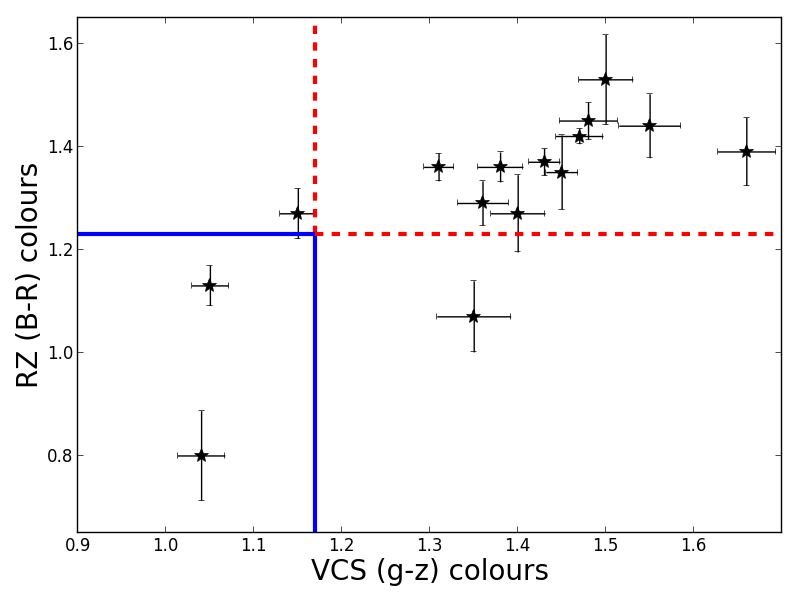}
\caption{The VCS versus RZ colours for the clusters that appear in both catalogues. The area enclosed by the broken (solid) lines show the area of parameter space where both the VCS and RZ are classifies as red (blue). The two sources for which the colours disagree are also shown.}
\label{colour_match}
\end{figure*}

For the VCS GC sample, the $g-z$ colour distribution is also bimodal, with peaks at roughly 1.1 and 1.45 and a dividing $g-z$ colour of approximately 1.17 \citep[see Fig.\,6 in ][]{2009ApJS..180...54J}. RZ01 found that their GC sample has a bimodal $B-R$ colour distribution with peaks at 1.1 and 1.35, with a dividing $B-R$ colour of 1.23. MKZ03 found that their GCs had a dividing colour of $V-I=1.10$, with blue and red peaks at 0.98 and 1.23 respectively.  We combined the three samples and applied both X-ray and optical completeness limits, where the 100\,\% optical completeness limits are $V\sim23.5$ (RZ01), $g=23.3$ and $z=22.2$ (VCS) and $V\sim23.5$ (MKZ03). This yielded a sample of 54 GCs hosting X-ray sources. We then found two GCs for which the colour assignments did not match across the three catalogues; these two clusters were not included in the analysis. The final completeness limited sample therefore consists of 37 red and 7 blue GCs. Figure\,\ref{colour_match} shows the VCS versus RZ colours for the clusters that appear in both catalogues. There is good agreement between the colours across the catalogues. The two sources for which there is a disagreement between the colours are also shown.

\par
The ratio of red to blue GCs hosting X-ray sources to be 5.3$\pm3.0$. When we account for the slightly larger number of blue GCs across the three samples, we find that ratio of red to blue GCs hosting X-ray ray sources to be 5.8$\pm2.8$. Previous studies \citep[see e.g.][]{2002ApJ...574L...5K,2003ApJ...586..814M,2004ApJ...613..279J,2007ApJ...660.1246S} have found that red GCs are roughly three times more likely to host a LMXB than blue GCs. \citet{2004ApJ...613..279J} provide the figure with the largest error, 3$\pm$1. 

\par
The mean 0.5--5\,keV luminosity for the red and blue subpopulations are 1.1$\pm1.2 \times10^{38}$\,erg\,s$^{-1}$ and 1.3$\pm0.4\times10^{38}$\,erg\,s$^{-1}$ respectively; they are statistically indistinguishable. However, using an AD test, we find the probability that the two X-ray luminosity distributions are drawn from the same sample to be only 4.9\,\%. Thus there is evidence that the red and blue GC-LMXB populations have different luminosity functions. This result is in contrast to the findings of \citet{2013ApJ...764...98K} and \citet{2016ApJ...818...33P} who found that the XLFs of red and blue subpopulations were similar across a large luminosity range.

\par
The radial distributions of the red and blue GC subpopulations were analysed. The mean radial distances for red and blue GC subpopulations were found to be 1.6$'\pm 0.8'$ and 2.0$'\pm 0.6'$ respectively. Using an AD test, we find that the probability of the two distributions being drawn from the same sample is 18\,\%. Thus there is no evidence to suggest that red and blue GC subpopulations have different radial distributions. 


\section{Conclusions}

We have found 238 X-ray point sources in the central 3.7$'$ of NGC\,4472; of which 191 were brighter than the 4\,$\sigma$ completeness limit of 6.3$\times10^{-16}$\,erg\,s$^{-1}$\,cm$^{-2}$ (0.5-2\,keV). The completeness limited XLF is well fit by a single power law model with slope and normalisation of 2.03$^{+0.16}_{-0.15}$ and $68.53^{+8.95}_{-8.28}$ respectively. Eighty of the 238 LMXBs in NGC\,4472 are found to reside in GCs. We have shown that there is no significant difference between the X-ray luminosities of the LMXBs in the field and those found in GCs. 

\par
Our study also shows that within the GC-LMXB population, the X-ray sources are preferentially found in red GCs at a ratio of roughly six to one. Our analysis of the red and blue GC subpopulations shows that they have indistinguishable radial distributions. However, there is evidence to suggest that the luminosity distributions of the red and blue GC-LMXB populations differ significantly, in contrast to other GC-LMXB systems. 

\section*{Acknowledgements}
TDJ acknowledges support from a Stobie-SALT studentship, funded jointly by the NRF of South Africa, the British Council and the University of Southampton. TDJ acknowledges a South Africa National Research Foundation Square Kilometre Array Research Fellowship. This material is based upon work supported by the National Aeronautics and Space Administration under Grants No. NAS8-03060 and GO1-12160X. This research has made use of data obtained from the Chandra Data Archive and software provided by the Chandra X-ray Center (CXC) in the application packages CIAO and ChIPS. Etc. TDJ would also like to thank Texas Tech University for its hospitality while this work was being finished and Mark Burke for his assistance.

\begin{table*}
\footnotesize
\label{source_list}

\caption[The table of the X-ray luminosities of X-ray sources in NGC\,4472]{ \footnotesize The sources within 3.7$'$ of the galaxy centre. The fluxes are in units of erg\,s$^{-1}$\,cm$^{-2}$ and the luminosities are in units of erg\,s$^{-1}$. The PL and DBB subscripts denote the spectral model conversion factor used to calculate the flux. The "err" designation denotes flux errors. The flag column identifies sources that are interesting and/ or for which caution must be used when analysing the source properties. The key is as follows: bkg = source in an area of high background, c = \emph{wavdetect} has misidentified or not detected the source in one of the observations, cg = the source is in or very close to a chip gap in one of the observations and GC = the source is associated with a globular cluster.} 

\begin{tabular}{|l|l|l|l|l|l|l|l|l|l|l|}
\hline
  \multicolumn{1}{|c|}{RA} &
  \multicolumn{1}{c|}{DEC} &
  \multicolumn{1}{c|}{Flux$_{\rm PL}$} &
  \multicolumn{1}{c|}{Flux$_{\rm PL}$ err} &
  \multicolumn{1}{c|}{Flux$_{\rm PL}$} &
  \multicolumn{1}{c|}{Flux$_{\rm PL}$ err} &
  \multicolumn{1}{c|}{Flux$_{\rm DBB}$} &
  \multicolumn{1}{c|}{Flux$_{\rm DBB}$ err} &
  \multicolumn{1}{c|}{Flux$_{\rm DBB}$} &
  \multicolumn{1}{c|}{Flux$_{\rm DBB}$ err} &
  \multicolumn{1}{c|}{Flag} \\
\midrule
\multicolumn{1}{|c|}{degrees} &
  \multicolumn{1}{c|}{degrees} &
  \multicolumn{1}{c|}{0.5--5\,keV} &
  \multicolumn{1}{c|}{0.5--5\,keV} &
  \multicolumn{1}{c|}{0.5--2\,keV} &
  \multicolumn{1}{c|}{0.5--2\,keV} &
  \multicolumn{1}{c|}{0.5--5\,keV} &
  \multicolumn{1}{c|}{0.5--5\,keV} &
  \multicolumn{1}{c|}{0.5--2\,keV} &
  \multicolumn{1}{c|}{0.5-2\,keV} & 
  \multicolumn{1}{c|}{} \\
\midrule
  187.49977 & 8.02063 & 1.46E-14 & 1.72E-15 & 7.64E-15 & 9.05E-16 & 1.54E-14 & 1.83E-15 & 6.25E-15 & 7.44E-16 & \\  
  187.49804 & 8.02989 & 9.14E-16 & 2.96E-16 & 4.8E-16 & 1.55E-16 & 9.64E-16 & 3.13E-16 & 3.92E-16 & 1.27E-16 & \\
  187.49757 & 7.99843 & 1.1E-15 & 2.59E-16 & 5.76E-16 & 1.36E-16 & 1.16E-15 & 2.73E-16 & 4.71E-16 & 1.11E-16 & \\
  187.49175 & 7.96768 & 2.96E-15 & 7.33E-16 & 1.55E-15 & 3.85E-16 & 3.12E-15 & 7.75E-16 & 1.27E-15 & 3.15E-16 & \\
  187.4915 & 7.96826 & 1.83E-15 & 3.88E-16 & 9.61E-16 & 2.04E-16 & 1.93E-15 & 4.1E-16 & 7.86E-16 & 1.67E-16 & \\
  187.49098 & 7.99121 & 1.57E-15 & 3.03E-16 & 8.24E-16 & 1.59E-16 & 1.66E-15 & 3.2E-16 & 6.74E-16 & 1.3E-16 & \\
  187.49021 & 7.9924 & 1.71E-15 & 3.07E-16 & 8.96E-16 & 1.61E-16 & 1.8E-15 & 3.25E-16 & 7.33E-16 & 1.32E-16 & GC\\
  187.4864 & 7.95771 & 1.17E-14 & 1.4E-15 & 6.16E-15 & 7.35E-16 & 1.24E-14 & 1.49E-15 & 5.04E-15 & 6.04E-16 & \\
  187.48501 & 8.00251 & 2.7E-15 & 4.04E-16 & 1.42E-15 & 2.12E-16 & 2.85E-15 & 4.28E-16 & 1.16E-15 & 1.74E-16 & \\
  187.48483 & 8.00848 & 4.12E-15 & 5.46E-16 & 2.16E-15 & 2.87E-16 & 4.35E-15 & 5.78E-16 & 1.77E-15 & 2.35E-16 & GC\\
  187.48422 & 8.01345 & 1.04E-15 & 2.82E-16 & 5.46E-16 & 1.48E-16 & 1.1E-15 & 2.97E-16 & 4.46E-16 & 1.21E-16 & \\
  187.48301 & 7.95625 & 2.27E-15 & 4.13E-16 & 1.19E-15 & 2.17E-16 & 2.4E-15 & 4.37E-16 & 9.76E-16 & 1.78E-16 & GC\\
  187.48287 & 8.00096 & 1.57E-15 & 2.99E-16 & 8.26E-16 & 1.57E-16 & 1.66E-15 & 3.16E-16 & 6.75E-16 & 1.29E-16 & GC\\
  187.48118 & 8.04989 & 1.09E-15 & 2.85E-16 & 5.74E-16 & 1.5E-16 & 1.15E-15 & 3.01E-16 & 4.69E-16 & 1.23E-16 & \\
  187.47628 & 8.01613 & 2.62E-15 & 3.81E-16 & 1.38E-15 & 2.0E-16 & 2.77E-15 & 4.04E-16 & 1.13E-15 & 1.64E-16 & \\
  187.47602 & 8.00271 & 4.86E-15 & 6.33E-16 & 2.55E-15 & 3.32E-16 & 5.12E-15 & 6.71E-16 & 2.08E-15 & 2.73E-16 & GC\\
  187.47566 & 7.95535 & 5.98E-15 & 7.76E-16 & 3.14E-15 & 4.08E-16 & 6.31E-15 & 8.23E-16 & 2.57E-15 & 3.35E-16 & \\
  187.47488 & 8.04315 & 1.48E-15 & 2.8E-16 & 7.74E-16 & 1.47E-16 & 1.56E-15 & 2.96E-16 & 6.33E-16 & 1.2E-16 & \\
  187.47378 & 7.99391 & 6.02E-15 & 7.74E-16 & 3.16E-15 & 4.06E-16 & 6.35E-15 & 8.2E-16 & 2.58E-15 & 3.34E-16 & \\
  187.47378 & 8.04469 & 3.28E-15 & 4.67E-16 & 1.72E-15 & 2.45E-16 & 3.46E-15 & 4.94E-16 & 1.41E-15 & 2.01E-16 & \\
  187.47234 & 8.00724 & 3.51E-15 & 4.78E-16 & 1.84E-15 & 2.51E-16 & 3.7E-15 & 5.06E-16 & 1.51E-15 & 2.06E-16 & GC\\
  187.47229 & 8.0526 & 3.17E-15 & 4.43E-16 & 1.67E-15 & 2.33E-16 & 3.35E-15 & 4.7E-16 & 1.36E-15 & 1.91E-16 & GC\\
  187.47228 & 8.01491 & 2.59E-15 & 3.8E-16 & 1.36E-15 & 1.99E-16 & 2.73E-15 & 4.02E-16 & 1.11E-15 & 1.63E-16 & \\
  187.47223 & 7.99779 & 3.61E-15 & 5.01E-16 & 1.9E-15 & 2.63E-16 & 3.81E-15 & 5.31E-16 & 1.55E-15 & 2.16E-16 & GC\\
  187.47144 & 7.95663 & 1.26E-15 & 3.09E-16 & 6.61E-16 & 1.62E-16 & 1.33E-15 & 3.27E-16 & 5.41E-16 & 1.33E-16 & \\
  187.47109 & 7.94715 & 1.18E-15 & 3.36E-16 & 6.17E-16 & 1.76E-16 & 1.24E-15 & 3.55E-16 & 5.05E-16 & 1.44E-16 & \\
  187.47044 & 7.99158 & 2.18E-15 & 4.97E-16 & 1.15E-15 & 2.61E-16 & 2.3E-15 & 5.26E-16 & 9.36E-16 & 2.14E-16 & \\
  187.47031 & 7.99906 & 4.21E-15 & 5.68E-16 & 2.21E-15 & 2.98E-16 & 4.44E-15 & 6.01E-16 & 1.81E-15 & 2.45E-16 & \\
  187.47021 & 7.96514 & 1.27E-15 & 3.67E-16 & 6.65E-16 & 1.93E-16 & 1.34E-15 & 3.87E-16 & 5.44E-16 & 1.58E-16 & \\
  187.47006 & 7.97202 & 2.93E-15 & 4.85E-16 & 1.54E-15 & 2.55E-16 & 3.09E-15 & 5.13E-16 & 1.26E-15 & 2.09E-16 & GC\\
  187.46972 & 8.03225 & 1.38E-14 & 1.62E-15 & 7.23E-15 & 8.48E-16 & 1.45E-14 & 1.71E-15 & 5.91E-15 & 6.97E-16 & \\
  187.46969 & 7.95261 & 1.78E-15 & 3.5E-16 & 9.35E-16 & 1.84E-16 & 1.88E-15 & 3.7E-16 & 7.64E-16 & 1.51E-16 & GC\\
  187.46921 & 8.01933 & 1.96E-15 & 3.07E-16 & 1.03E-15 & 1.61E-16 & 2.06E-15 & 3.25E-16 & 8.39E-16 & 1.32E-16 & GC\\
  187.46914 & 8.00449 & 1.65E-15 & 3.61E-16 & 8.67E-16 & 1.89E-16 & 1.74E-15 & 3.81E-16 & 7.09E-16 & 1.55E-16 & GC\\
  187.46883 & 8.00051 & 3.14E-15 & 4.58E-16 & 1.65E-15 & 2.41E-16 & 3.31E-15 & 4.85E-16 & 1.35E-15 & 1.97E-16 & GC\\
  187.46816 & 7.99314 & 3.26E-15 & 9.31E-16 & 1.71E-15 & 4.89E-16 & 3.43E-15 & 9.83E-16 & 1.4E-15 & 4.0E-16 & \\
  187.46788 & 8.01287 & 8.14E-16 & 2.2E-16 & 4.27E-16 & 1.15E-16 & 8.59E-16 & 2.32E-16 & 3.49E-16 & 9.44E-17 & \\
  187.46582 & 7.99676 & 3.14E-15 & 9.44E-16 & 1.65E-15 & 4.96E-16 & 3.31E-15 & 9.97E-16 & 1.35E-15 & 4.06E-16 & GC\\
  187.46582 & 8.03906 & 1.48E-15 & 3.38E-16 & 7.79E-16 & 1.77E-16 & 1.57E-15 & 3.57E-16 & 6.37E-16 & 1.45E-16 & cg\\
  187.46556 & 7.99996 & 4.17E-15 & 5.69E-16 & 2.19E-15 & 2.99E-16 & 4.4E-15 & 6.03E-16 & 1.79E-15 & 2.45E-16 & GC\\
  187.46548 & 7.9732 & 2.56E-15 & 4.07E-16 & 1.34E-15 & 2.13E-16 & 2.7E-15 & 4.3E-16 & 1.1E-15 & 1.75E-16 & \\
  187.46543 & 7.99196 & 1.39E-15 & 3.23E-16 & 7.27E-16 & 1.7E-16 & 1.46E-15 & 3.41E-16 & 5.95E-16 & 1.39E-16 & GC\\
  187.46507 & 8.00083 & 3.99E-15 & 1.05E-15 & 2.1E-15 & 5.5E-16 & 4.21E-15 & 1.11E-15 & 1.71E-15 & 4.5E-16 & \\
  187.46489 & 7.96636 & 2.55E-15 & 4.6E-16 & 1.34E-15 & 2.42E-16 & 2.69E-15 & 4.87E-16 & 1.09E-15 & 1.98E-16 & \\
  187.46478 & 7.95126 & 2.51E-15 & 4.85E-16 & 1.32E-15 & 2.55E-16 & 2.65E-15 & 5.13E-16 & 1.08E-15 & 2.09E-16 & \\
  187.46394 & 7.98809 & 2.08E-15 & 3.7E-16 & 1.09E-15 & 1.94E-16 & 2.19E-15 & 3.91E-16 & 8.91E-16 & 1.59E-16 & GC\\
  187.4634 & 8.05763 & 3.52E-15 & 9.38E-16 & 1.85E-15 & 4.92E-16 & 3.72E-15 & 9.9E-16 & 1.51E-15 & 4.03E-16 & \\
  187.46323 & 7.97408 & 1.04E-15 & 2.63E-16 & 5.44E-16 & 1.38E-16 & 1.09E-15 & 2.78E-16 & 4.45E-16 & 1.13E-16 & \\
  187.46315 & 8.04035 & 7.28E-15 & 9.14E-16 & 3.82E-15 & 4.8E-16 & 7.68E-15 & 9.69E-16 & 3.12E-15 & 3.94E-16 & \\
  187.46276 & 7.99645 & 3.37E-15 & 1.04E-15 & 1.77E-15 & 5.46E-16 & 3.56E-15 & 1.1E-15 & 1.45E-15 & 4.47E-16 & GC\\
  187.46225 & 7.98736 & 1.49E-15 & 3.46E-16 & 7.79E-16 & 1.82E-16 & 1.57E-15 & 3.66E-16 & 6.37E-16 & 1.49E-16 & GC\\
  187.462 & 8.00271 & 9.52E-15 & 1.14E-15 & 5.0E-15 & 6.01E-16 & 1.0E-14 & 1.21E-15 & 4.08E-15 & 4.93E-16 & GC\\
  187.46196 & 8.03652 & 2.07E-15 & 3.6E-16 & 1.09E-15 & 1.89E-16 & 2.18E-15 & 3.81E-16 & 8.87E-16 & 1.55E-16 & \\
  187.46183 & 7.9715 & 1.95E-15 & 3.68E-16 & 1.02E-15 & 1.93E-16 & 2.05E-15 & 3.89E-16 & 8.35E-16 & 1.58E-16 & \\
  187.46169 & 8.04609 & 1.04E-15 & 1.99E-16 & 5.44E-16 & 1.05E-16 & 1.09E-15 & 2.11E-16 & 4.45E-16 & 8.58E-17 & \\
  187.46132 & 8.01972 & 1.18E-15 & 2.38E-16 & 6.2E-16 & 1.25E-16 & 1.25E-15 & 2.51E-16 & 5.07E-16 & 1.02E-16 & GC\\
  187.46106 & 7.96407 & 2.05E-15 & 4.9E-16 & 1.07E-15 & 2.57E-16 & 2.16E-15 & 5.18E-16 & 8.77E-16 & 2.11E-16 & \\
  187.46101 & 8.00652 & 6.63E-16 & 2.4E-16 & 3.48E-16 & 1.26E-16 & 6.99E-16 & 2.53E-16 & 2.84E-16 & 1.03E-16 & \\
  187.46063 & 7.98888 & 3.41E-15 & 4.96E-16 & 1.79E-15 & 2.6E-16 & 3.6E-15 & 5.25E-16 & 1.47E-15 & 2.13E-16 & \\
  
\end{tabular}
\end{table*}

\begin{table*}
\footnotesize
\label{source_list}

\contcaption{ \footnotesize The sources within 3.7$'$ of the galaxy centre. The fluxes are in units of erg\,s$^{-1}$\,cm$^{-2}$ and the luminosities are in units of erg\,s$^{-1}$. The PL and DBB subscripts denote the spectral model conversion factor used to calculate the flux. The "err" designation denotes flux errors. The flag column identifies sources that are interesting and/ or for which caution must be used when analysing the source properties. The key is as follows: bkg = source in an area of high background, c = \emph{wavdetect} has misidentified or not detected the source in one of the observations, cg = the source is in or very close to a chip gap in one of the observations and GC = the source is associated with a globular cluster.} 

\begin{tabular}{|l|l|l|l|l|l|l|l|l|l|l|}
\hline
  \multicolumn{1}{|c|}{RA} &
  \multicolumn{1}{c|}{DEC} &
  \multicolumn{1}{c|}{Flux$_{\rm PL}$} &
  \multicolumn{1}{c|}{Flux$_{\rm PL}$ err} &
  \multicolumn{1}{c|}{Flux$_{\rm PL}$} &
  \multicolumn{1}{c|}{Flux$_{\rm PL}$ err} &
  \multicolumn{1}{c|}{Flux$_{\rm DBB}$} &
  \multicolumn{1}{c|}{Flux$_{\rm DBB}$ err} &
  \multicolumn{1}{c|}{Flux$_{\rm DBB}$} &
  \multicolumn{1}{c|}{Flux$_{\rm DBB}$ err} &
  \multicolumn{1}{c|}{Flag} \\
\midrule
\multicolumn{1}{|c|}{degrees} &
  \multicolumn{1}{c|}{degrees} &
  \multicolumn{1}{c|}{0.5--5\,keV} &
  \multicolumn{1}{c|}{0.5--5\,keV} &
  \multicolumn{1}{c|}{0.5--2\,keV} &
  \multicolumn{1}{c|}{0.5--2\,keV} &
  \multicolumn{1}{c|}{0.5--5\,keV} &
  \multicolumn{1}{c|}{0.5--5\,keV} &
  \multicolumn{1}{c|}{0.5--2\,keV} &
  \multicolumn{1}{c|}{0.5-2\,keV} & 
  \multicolumn{1}{c|}{} \\
\midrule

  187.46004 & 8.00379 & 9.21E-15 & 1.11E-15 & 4.83E-15 & 5.82E-16 & 9.72E-15 & 1.18E-15 & 3.95E-15 & 4.78E-16 & GC\\
  187.45957 & 8.00999 & 1.86E-15 & 3.42E-16 & 9.78E-16 & 1.79E-16 & 1.97E-15 & 3.61E-16 & 8.0E-16 & 1.47E-16 & \\
  187.4594 & 7.97335 & 1.96E-15 & 3.71E-16 & 1.03E-15 & 1.95E-16 & 2.06E-15 & 3.93E-16 & 8.4E-16 & 1.6E-16 & \\
  187.45886 & 8.02188 & 1.45E-15 & 2.75E-16 & 7.63E-16 & 1.45E-16 & 1.53E-15 & 2.91E-16 & 6.24E-16 & 1.18E-16 & \\
  187.45878 & 7.99547 & 7.63E-15 & 9.79E-16 & 4.0E-15 & 5.14E-16 & 8.04E-15 & 1.04E-15 & 3.27E-15 & 4.22E-16 & GC\\
  187.45789 & 7.99961 & 1.86E-15 & 5.43E-16 & 9.79E-16 & 2.85E-16 & 1.97E-15 & 5.73E-16 & 8.0E-16 & 2.33E-16 & bkg\\
  187.45758 & 8.00714 & 1.07E-15 & 2.57E-16 & 5.6E-16 & 1.35E-16 & 1.13E-15 & 2.71E-16 & 4.58E-16 & 1.1E-16 & \\
  187.45639 & 7.98617 & 1.28E-15 & 3.31E-16 & 6.74E-16 & 1.74E-16 & 1.36E-15 & 3.5E-16 & 5.51E-16 & 1.42E-16 & \\
  187.45636 & 7.99033 & 1.62E-15 & 4.87E-16 & 8.48E-16 & 2.56E-16 & 1.7E-15 & 5.14E-16 & 6.93E-16 & 2.09E-16 & bkg, GC\\
  187.45612 & 7.99293 & 1.86E-15 & 4.04E-16 & 9.78E-16 & 2.12E-16 & 1.97E-15 & 4.27E-16 & 8.0E-16 & 1.74E-16 & GC\\
  187.45608 & 8.05566 & 8.28E-16 & 1.87E-16 & 4.34E-16 & 9.79E-17 & 8.73E-16 & 1.97E-16 & 3.55E-16 & 8.02E-17 & \\
  187.45568 & 7.96479 & 5.58E-15 & 7.49E-16 & 2.93E-15 & 3.93E-16 & 5.88E-15 & 7.94E-16 & 2.39E-15 & 3.23E-16 & GC\\
  187.45552 & 8.00358 & 1.52E-14 & 1.79E-15 & 7.98E-15 & 9.39E-16 & 1.6E-14 & 1.9E-15 & 6.52E-15 & 7.71E-16 & \\
  187.4555 & 8.03852 & 1.23E-15 & 2.73E-16 & 6.46E-16 & 1.43E-16 & 1.3E-15 & 2.89E-16 & 5.28E-16 & 1.17E-16 & GC\\
  187.45524 & 8.02964 & 1.96E-15 & 3.12E-16 & 1.03E-15 & 1.64E-16 & 2.07E-15 & 3.3E-16 & 8.41E-16 & 1.34E-16 & \\
  187.45509 & 8.05476 & 1.24E-15 & 3.06E-16 & 6.52E-16 & 1.61E-16 & 1.31E-15 & 3.23E-16 & 5.33E-16 & 1.31E-16 & c\\
  187.45507 & 8.0155 & 7.23E-16 & 2.62E-16 & 3.8E-16 & 1.38E-16 & 7.63E-16 & 2.77E-16 & 3.1E-16 & 1.13E-16 & GC\\
  187.45461 & 8.0076 & 8.41E-16 & 2.18E-16 & 4.41E-16 & 1.14E-16 & 8.87E-16 & 2.3E-16 & 3.61E-16 & 9.36E-17 & GC\\
  187.45438 & 7.99198 & 1.52E-15 & 4.46E-16 & 7.95E-16 & 2.34E-16 & 1.6E-15 & 4.7E-16 & 6.5E-16 & 1.91E-16 & \\
  187.45426 & 7.98095 & 1.5E-14 & 1.76E-15 & 7.88E-15 & 9.26E-16 & 1.58E-14 & 1.87E-15 & 6.44E-15 & 7.61E-16 & \\
  187.45416 & 7.99612 & 2.81E-15 & 4.66E-16 & 1.48E-15 & 2.45E-16 & 2.97E-15 & 4.93E-16 & 1.21E-15 & 2.01E-16 & \\
  187.45379 & 7.97948 & 1.73E-15 & 4.34E-16 & 9.07E-16 & 2.28E-16 & 1.82E-15 & 4.59E-16 & 7.41E-16 & 1.87E-16 & GC\\
  187.45375 & 8.01616 & 4.87E-16 & 2.1E-16 & 2.56E-16 & 1.1E-16 & 5.14E-16 & 2.21E-16 & 2.09E-16 & 9.0E-17 & \\
  187.45308 & 8.02145 & 1.11E-15 & 3.36E-16 & 5.84E-16 & 1.76E-16 & 1.18E-15 & 3.55E-16 & 4.78E-16 & 1.44E-16 & \\
  187.45261 & 7.9984 & 2.16E-15 & 4.14E-16 & 1.13E-15 & 2.17E-16 & 2.28E-15 & 4.38E-16 & 9.26E-16 & 1.78E-16 & \\
  187.45261 & 7.98706 & 1.9E-15 & 4.38E-16 & 9.98E-16 & 2.3E-16 & 2.01E-15 & 4.62E-16 & 8.16E-16 & 1.88E-16 & \\
  187.4525 & 7.99542 & 5.88E-15 & 8.37E-16 & 3.09E-15 & 4.4E-16 & 6.2E-15 & 8.87E-16 & 2.52E-15 & 3.61E-16 & \\
  187.45198 & 8.00991 & 1.55E-15 & 4.8E-16 & 8.15E-16 & 2.52E-16 & 1.64E-15 & 5.07E-16 & 6.66E-16 & 2.06E-16 & bkg\\
  187.45187 & 8.03341 & 9.42E-16 & 2.0E-16 & 4.94E-16 & 1.05E-16 & 9.93E-16 & 2.11E-16 & 4.04E-16 & 8.58E-17 & \\
  187.45155 & 8.00554 & 8.56E-16 & 3.34E-16 & 4.49E-16 & 1.75E-16 & 9.03E-16 & 3.52E-16 & 3.67E-16 & 1.43E-16 & \\
  187.45135 & 7.99231 & 3.51E-15 & 5.81E-16 & 1.84E-15 & 3.05E-16 & 3.7E-15 & 6.15E-16 & 1.51E-15 & 2.5E-16 & \\
  187.45113 & 8.00454 & 2.89E-15 & 1.09E-15 & 1.52E-15 & 5.74E-16 & 3.05E-15 & 1.15E-15 & 1.24E-15 & 4.69E-16 & \\
  187.45073 & 8.01639 & 1.49E-15 & 2.71E-16 & 7.79E-16 & 1.42E-16 & 1.57E-15 & 2.86E-16 & 6.37E-16 & 1.17E-16 & \\
  187.45034 & 7.99476 & 1.44E-15 & 4.38E-16 & 7.54E-16 & 2.3E-16 & 1.52E-15 & 4.62E-16 & 6.17E-16 & 1.88E-16 & GC\\
  187.45009 & 7.97628 & 6.51E-16 & 2.47E-16 & 3.42E-16 & 1.3E-16 & 6.87E-16 & 2.61E-16 & 2.8E-16 & 1.06E-16 & GC\\
  187.45006 & 7.98558 & 1.54E-15 & 3.65E-16 & 8.1E-16 & 1.92E-16 & 1.63E-15 & 3.86E-16 & 6.62E-16 & 1.57E-16 & \\
  187.44982 & 8.00233 & 8.39E-15 & 1.36E-15 & 4.4E-15 & 7.14E-16 & 8.85E-15 & 1.44E-15 & 3.6E-15 & 5.85E-16 & bkg, GC\\
  187.44947 & 7.99776 & 3.19E-15 & 7.79E-16 & 1.67E-15 & 4.09E-16 & 3.36E-15 & 8.23E-16 & 1.37E-15 & 3.35E-16 & bkg\\
  187.44946 & 8.01021 & 3.44E-15 & 5.42E-16 & 1.81E-15 & 2.84E-16 & 3.63E-15 & 5.73E-16 & 1.48E-15 & 2.33E-16 & \\
  187.44942 & 7.99556 & 1.32E-15 & 4.35E-16 & 6.91E-16 & 2.28E-16 & 1.39E-15 & 4.59E-16 & 5.65E-16 & 1.87E-16 & \\
  187.44941 & 7.98863 & 2.18E-15 & 4.32E-16 & 1.15E-15 & 2.27E-16 & 2.3E-15 & 4.57E-16 & 9.36E-16 & 1.86E-16 & GC\\
  187.44936 & 7.99283 & 2.55E-15 & 4.88E-16 & 1.34E-15 & 2.56E-16 & 2.69E-15 & 5.16E-16 & 1.09E-15 & 2.1E-16 & \\
  187.4492 & 8.00554 & 1.08E-15 & 4.19E-16 & 5.67E-16 & 2.2E-16 & 1.14E-15 & 4.43E-16 & 4.63E-16 & 1.8E-16 & bkg\\
  187.44894 & 7.99055 & 4.14E-15 & 5.99E-16 & 2.17E-15 & 3.14E-16 & 4.37E-15 & 6.34E-16 & 1.78E-15 & 2.58E-16 & GC\\
  187.44881 & 7.98298 & 1.43E-15 & 3.82E-16 & 7.51E-16 & 2.0E-16 & 1.51E-15 & 4.03E-16 & 6.14E-16 & 1.64E-16 & \\
  187.44873 & 8.02852 & 1.28E-15 & 2.5E-16 & 6.74E-16 & 1.31E-16 & 1.35E-15 & 2.65E-16 & 5.51E-16 & 1.08E-16 & \\
  187.44868 & 8.02199 & 1.49E-15 & 2.96E-16 & 7.8E-16 & 1.55E-16 & 1.57E-15 & 3.13E-16 & 6.38E-16 & 1.27E-16 & \\
  187.44867 & 8.0074 & 7.79E-15 & 9.76E-16 & 4.09E-15 & 5.12E-16 & 8.22E-15 & 1.04E-15 & 3.34E-15 & 4.21E-16 & \\
  187.44849 & 7.97868 & 1.06E-15 & 2.77E-16 & 5.54E-16 & 1.45E-16 & 1.11E-15 & 2.92E-16 & 4.53E-16 & 1.19E-16 & \\
  187.44765 & 8.00222 & 1.47E-14 & 2.0E-15 & 7.71E-15 & 1.05E-15 & 1.55E-14 & 2.12E-15 & 6.3E-15 & 8.61E-16 & \\
  187.44753 & 7.98658 & 1.1E-15 & 3.52E-16 & 5.78E-16 & 1.85E-16 & 1.16E-15 & 3.71E-16 & 4.72E-16 & 1.51E-16 & GC\\
  187.44716 & 8.01575 & 1.04E-15 & 2.41E-16 & 5.43E-16 & 1.26E-16 & 1.09E-15 & 2.54E-16 & 4.44E-16 & 1.04E-16 & GC\\
  187.4465 & 8.038 & 7.06E-16 & 2.27E-16 & 3.7E-16 & 1.19E-16 & 7.44E-16 & 2.4E-16 & 3.03E-16 & 9.76E-17 & \\
  187.44648 & 7.97605 & 1.78E-15 & 3.49E-16 & 9.32E-16 & 1.83E-16 & 1.87E-15 & 3.68E-16 & 7.62E-16 & 1.5E-16 & GC\\
  187.44604 & 8.03101 & 7.59E-16 & 1.83E-16 & 3.98E-16 & 9.63E-17 & 8.01E-16 & 1.94E-16 & 3.26E-16 & 7.88E-17 & cg\\
  187.44602 & 8.00373 & 1.91E-14 & 2.28E-15 & 1.0E-14 & 1.2E-15 & 2.01E-14 & 2.42E-15 & 8.19E-15 & 9.83E-16 & \\
  187.44598 & 7.99539 & 1.24E-15 & 4.57E-16 & 6.53E-16 & 2.4E-16 & 1.31E-15 & 4.83E-16 & 5.34E-16 & 1.96E-16 & \\
  187.44595 & 7.97969 & 5.78E-15 & 7.48E-16 & 3.03E-15 & 3.93E-16 & 6.1E-15 & 7.93E-16 & 2.48E-15 & 3.22E-16 & GC\\
 
\end{tabular}
\end{table*}

\begin{table*}
\footnotesize
\label{source_list}

\contcaption{ \footnotesize The sources within 3.7$'$ of the galaxy centre. The fluxes are in units of erg\,s$^{-1}$\,cm$^{-2}$ and the luminosities are in units of erg\,s$^{-1}$. The PL and DBB subscripts denote the spectral model conversion factor used to calculate the flux. The "err" designation denotes flux errors. The flag column identifies sources that are interesting and/ or for which caution must be used when analysing the source properties. The key is as follows: bkg = source in an area of high background, c = \emph{wavdetect} has misidentified or not detected the source in one of the observations, cg = the source is in or very close to a chip gap in one of the observations and GC = the source is associated with a globular cluster.} 

\begin{tabular}{|l|l|l|l|l|l|l|l|l|l|l|}
\hline
  \multicolumn{1}{|c|}{RA} &
  \multicolumn{1}{c|}{DEC} &
  \multicolumn{1}{c|}{Flux$_{\rm PL}$} &
  \multicolumn{1}{c|}{Flux$_{\rm PL}$ err} &
  \multicolumn{1}{c|}{Flux$_{\rm PL}$} &
  \multicolumn{1}{c|}{Flux$_{\rm PL}$ err} &
  \multicolumn{1}{c|}{Flux$_{\rm DBB}$} &
  \multicolumn{1}{c|}{Flux$_{\rm DBB}$ err} &
  \multicolumn{1}{c|}{Flux$_{\rm DBB}$} &
  \multicolumn{1}{c|}{Flux$_{\rm DBB}$ err} &
  \multicolumn{1}{c|}{Flag} \\
\midrule
\multicolumn{1}{|c|}{degrees} &
  \multicolumn{1}{c|}{degrees} &
  \multicolumn{1}{c|}{0.5--5\,keV} &
  \multicolumn{1}{c|}{0.5--5\,keV} &
  \multicolumn{1}{c|}{0.5--2\,keV} &
  \multicolumn{1}{c|}{0.5--2\,keV} &
  \multicolumn{1}{c|}{0.5--5\,keV} &
  \multicolumn{1}{c|}{0.5--5\,keV} &
  \multicolumn{1}{c|}{0.5--2\,keV} &
  \multicolumn{1}{c|}{0.5-2\,keV} & 
  \multicolumn{1}{c|}{} \\
\midrule
  187.44517 & 7.98295 & 1.7E-15 & 3.64E-16 & 8.91E-16 & 1.91E-16 & 1.79E-15 & 3.85E-16 & 7.29E-16 & 1.57E-16 & \\
  187.44418 & 7.99237 & 2.18E-15 & 4.25E-16 & 1.15E-15 & 2.23E-16 & 2.3E-15 & 4.49E-16 & 9.37E-16 & 1.83E-16 & \\
  187.44399 & 7.99015 & 1.97E-15 & 3.85E-16 & 1.04E-15 & 2.02E-16 & 2.08E-15 & 4.06E-16 & 8.46E-16 & 1.65E-16 & \\
  187.44364 & 7.99437 & 1.6E-15 & 4.42E-16 & 8.38E-16 & 2.32E-16 & 1.68E-15 & 4.66E-16 & 6.85E-16 & 1.9E-16 & \\
  187.44349 & 8.01867 & 1.4E-15 & 3.69E-16 & 7.35E-16 & 1.94E-16 & 1.48E-15 & 3.9E-16 & 6.01E-16 & 1.59E-16 & \\
  187.44335 & 8.02061 & 9.58E-16 & 2.93E-16 & 5.03E-16 & 1.54E-16 & 1.01E-15 & 3.09E-16 & 4.11E-16 & 1.26E-16 & \\
  187.44327 & 7.96177 & 1.51E-15 & 3.77E-16 & 7.94E-16 & 1.98E-16 & 1.6E-15 & 3.98E-16 & 6.49E-16 & 1.62E-16 & \\
  187.44317 & 8.02419 & 3.13E-15 & 4.35E-16 & 1.64E-15 & 2.28E-16 & 3.3E-15 & 4.6E-16 & 1.34E-15 & 1.87E-16 & GC\\
  187.44288 & 7.97872 & 2.15E-15 & 3.95E-16 & 1.13E-15 & 2.07E-16 & 2.26E-15 & 4.17E-16 & 9.2E-16 & 1.7E-16 & \\
  187.44288 & 8.02239 & 9.69E-16 & 2.28E-16 & 5.08E-16 & 1.2E-16 & 1.02E-15 & 2.41E-16 & 4.16E-16 & 9.8E-17 & \\
  187.4425 & 7.99815 & 2.88E-14 & 3.44E-15 & 1.51E-14 & 1.8E-15 & 3.04E-14 & 3.64E-15 & 1.23E-14 & 1.48E-15 & \\
  187.44231 & 7.98961 & 3.31E-15 & 5.06E-16 & 1.74E-15 & 2.66E-16 & 3.49E-15 & 5.35E-16 & 1.42E-15 & 2.18E-16 & GC\\
  187.44205 & 8.02309 & 1.15E-15 & 2.42E-16 & 6.04E-16 & 1.27E-16 & 1.21E-15 & 2.56E-16 & 4.94E-16 & 1.04E-16 & GC\\
  187.44202 & 7.98761 & 1.38E-14 & 1.64E-15 & 7.24E-15 & 8.62E-16 & 1.46E-14 & 1.74E-15 & 5.92E-15 & 7.08E-16 & GC\\
  187.44131 & 8.02414 & 5.76E-15 & 7.22E-16 & 3.02E-15 & 3.79E-16 & 6.08E-15 & 7.66E-16 & 2.47E-15 & 3.11E-16 & GC\\
  187.44093 & 8.00475 & 4.39E-15 & 8.92E-16 & 2.3E-15 & 4.68E-16 & 4.63E-15 & 9.43E-16 & 1.88E-15 & 3.84E-16 & \\
  187.44087 & 7.98851 & 2.84E-15 & 5.14E-16 & 1.49E-15 & 2.7E-16 & 3.0E-15 & 5.43E-16 & 1.22E-15 & 2.21E-16 & \\
  187.44078 & 7.96204 & 2.82E-15 & 5.15E-16 & 1.48E-15 & 2.7E-16 & 2.98E-15 & 5.44E-16 & 1.21E-15 & 2.21E-16 & GC\\
  187.4406 & 7.9692 & 3.14E-15 & 4.77E-16 & 1.65E-15 & 2.51E-16 & 3.31E-15 & 5.05E-16 & 1.35E-15 & 2.05E-16 & \\
  187.44054 & 8.02476 & 2.58E-15 & 3.77E-16 & 1.35E-15 & 1.98E-16 & 2.72E-15 & 3.99E-16 & 1.11E-15 & 1.62E-16 & c\\
  187.44023 & 7.99462 & 1.04E-15 & 3.31E-16 & 5.44E-16 & 1.74E-16 & 1.09E-15 & 3.49E-16 & 4.45E-16 & 1.42E-16 & GC\\
  187.44004 & 7.99341 & 7.4E-16 & 2.68E-16 & 3.89E-16 & 1.41E-16 & 7.81E-16 & 2.83E-16 & 3.18E-16 & 1.15E-16 & GC\\
  187.43988 & 8.02479 & 2.35E-15 & 3.58E-16 & 1.24E-15 & 1.88E-16 & 2.48E-15 & 3.79E-16 & 1.01E-15 & 1.54E-16 & c\\
  187.43985 & 8.0164 & 6.02E-15 & 7.61E-16 & 3.16E-15 & 4.0E-16 & 6.35E-15 & 8.07E-16 & 2.58E-15 & 3.28E-16 & \\
  187.43968 & 8.03374 & 3.11E-15 & 4.28E-16 & 1.63E-15 & 2.25E-16 & 3.28E-15 & 4.54E-16 & 1.33E-15 & 1.85E-16 & \\
  187.43941 & 7.99174 & 2.56E-15 & 4.79E-16 & 1.34E-15 & 2.52E-16 & 2.7E-15 & 5.07E-16 & 1.1E-15 & 2.06E-16 & \\
  187.4392 & 7.98576 & 2.35E-15 & 5.29E-16 & 1.23E-15 & 2.78E-16 & 2.48E-15 & 5.59E-16 & 1.01E-15 & 2.27E-16 & \\
  187.43919 & 7.98781 & 1.87E-15 & 4.26E-16 & 9.81E-16 & 2.24E-16 & 1.97E-15 & 4.5E-16 & 8.02E-16 & 1.83E-16 & \\
  187.4391 & 7.99342 & 1.71E-15 & 3.85E-16 & 8.98E-16 & 2.02E-16 & 1.81E-15 & 4.07E-16 & 7.35E-16 & 1.66E-16 & GC\\
  187.439 & 8.0046 & 7.38E-15 & 9.98E-16 & 3.87E-15 & 5.24E-16 & 7.78E-15 & 1.06E-15 & 3.17E-15 & 4.3E-16 & \\
  187.43899 & 7.99449 & 2.04E-15 & 4.14E-16 & 1.07E-15 & 2.17E-16 & 2.15E-15 & 4.37E-16 & 8.74E-16 & 1.78E-16 & GC\\
  187.43893 & 8.00817 & 1.99E-15 & 3.85E-16 & 1.04E-15 & 2.02E-16 & 2.1E-15 & 4.07E-16 & 8.52E-16 & 1.65E-16 & \\
  187.43846 & 8.02502 & 4.56E-16 & 1.96E-16 & 2.39E-16 & 1.03E-16 & 4.81E-16 & 2.07E-16 & 1.96E-16 & 8.42E-17 & \\
  187.43825 & 7.94748 & 1.69E-15 & 4.53E-16 & 8.87E-16 & 2.38E-16 & 1.78E-15 & 4.79E-16 & 7.25E-16 & 1.95E-16 & \\
  187.43823 & 8.01061 & 4.85E-16 & 2.33E-16 & 2.55E-16 & 1.22E-16 & 5.12E-16 & 2.46E-16 & 2.08E-16 & 1.0E-16 & bkg\\
  187.43763 & 7.99741 & 3.48E-15 & 5.13E-16 & 1.82E-15 & 2.69E-16 & 3.67E-15 & 5.43E-16 & 1.49E-15 & 2.21E-16 & GC\\
  187.43755 & 8.03682 & 1.95E-15 & 3.24E-16 & 1.03E-15 & 1.7E-16 & 2.06E-15 & 3.42E-16 & 8.38E-16 & 1.39E-16 & \\
  187.43733 & 8.01012 & 9.53E-16 & 3.29E-16 & 5.0E-16 & 1.73E-16 & 1.01E-15 & 3.47E-16 & 4.09E-16 & 1.41E-16 & bkg\\
  187.4368 & 8.0013 & 1.96E-15 & 3.7E-16 & 1.03E-15 & 1.94E-16 & 2.07E-15 & 3.91E-16 & 8.41E-16 & 1.59E-16 & \\
  187.43675 & 8.05909 & 1.08E-15 & 2.89E-16 & 5.68E-16 & 1.52E-16 & 1.14E-15 & 3.05E-16 & 4.64E-16 & 1.24E-16 & \\
  187.4361 & 8.00019 & 4.43E-15 & 6.12E-16 & 2.33E-15 & 3.21E-16 & 4.68E-15 & 6.48E-16 & 1.9E-15 & 2.64E-16 & \\
  187.43583 & 7.95621 & 2.79E-15 & 5.13E-16 & 1.46E-15 & 2.69E-16 & 2.94E-15 & 5.42E-16 & 1.2E-15 & 2.21E-16 & \\
  187.43494 & 8.03965 & 9.25E-16 & 3.23E-16 & 4.86E-16 & 1.7E-16 & 9.76E-16 & 3.41E-16 & 3.97E-16 & 1.39E-16 & \\
  187.43483 & 7.99499 & 9.4E-16 & 3.1E-16 & 4.93E-16 & 1.63E-16 & 9.92E-16 & 3.27E-16 & 4.03E-16 & 1.33E-16 & GC\\
  187.43415 & 8.01161 & 3.25E-15 & 4.72E-16 & 1.71E-15 & 2.48E-16 & 3.43E-15 & 5.0E-16 & 1.4E-15 & 2.03E-16 & GC\\
  187.43395 & 8.05697 & 1.14E-15 & 3.08E-16 & 5.99E-16 & 1.62E-16 & 1.2E-15 & 3.26E-16 & 4.9E-16 & 1.32E-16 & \\
  187.43375 & 7.97804 & 6.34E-15 & 8.32E-16 & 3.33E-15 & 4.37E-16 & 6.69E-15 & 8.82E-16 & 2.72E-15 & 3.59E-16 & GC\\
  187.43343 & 8.02091 & 1.92E-15 & 6.93E-16 & 1.01E-15 & 3.64E-16 & 2.03E-15 & 7.31E-16 & 8.24E-16 & 2.98E-16 & GC\\
  187.43332 & 7.99764 & 5.77E-15 & 7.61E-16 & 3.03E-15 & 4.0E-16 & 6.09E-15 & 8.06E-16 & 2.48E-15 & 3.28E-16 & GC\\
  187.43302 & 8.04708 & 9.81E-15 & 1.74E-15 & 5.15E-15 & 9.14E-16 & 1.04E-14 & 1.84E-15 & 4.21E-15 & 7.49E-16 & \\
  187.43289 & 8.05309 & 6.23E-16 & 2.13E-16 & 3.27E-16 & 1.12E-16 & 6.57E-16 & 2.24E-16 & 2.67E-16 & 9.13E-17 & \\
  187.43269 & 7.96569 & 3.06E-15 & 6.56E-16 & 1.6E-15 & 3.44E-16 & 3.22E-15 & 6.93E-16 & 1.31E-15 & 2.82E-16 & GC\\
  187.43259 & 7.99911 & 1.52E-15 & 3.26E-16 & 8.0E-16 & 1.71E-16 & 1.61E-15 & 3.45E-16 & 6.54E-16 & 1.4E-16 & GC\\
  187.43248 & 7.9543 & 4.08E-15 & 5.94E-16 & 2.14E-15 & 3.12E-16 & 4.31E-15 & 6.29E-16 & 1.75E-15 & 2.56E-16 & \\
  187.43245 & 7.95842 & 1.88E-15 & 4.22E-16 & 9.89E-16 & 2.21E-16 & 1.99E-15 & 4.46E-16 & 8.08E-16 & 1.81E-16 & \\
  187.43237 & 8.01162 & 2.42E-15 & 4.0E-16 & 1.27E-15 & 2.1E-16 & 2.56E-15 & 4.23E-16 & 1.04E-15 & 1.72E-16 & GC\\
  187.43185 & 8.01442 & 3.08E-15 & 4.39E-16 & 1.62E-15 & 2.31E-16 & 3.25E-15 & 4.65E-16 & 1.32E-15 & 1.89E-16 & \\
  187.43142 & 7.98188 & 1.25E-15 & 3.45E-16 & 6.57E-16 & 1.81E-16 & 1.32E-15 & 3.65E-16 & 5.37E-16 & 1.48E-16 & \\
 187.43135 & 7.99651 & 9.21E-15 & 1.13E-15 & 4.83E-15 & 5.95E-16 & 9.71E-15 & 1.2E-15 & 3.95E-15 & 4.88E-16 & \\
   
  \end{tabular}
\end{table*}
\begin{table*}
\footnotesize
\label{source_list}

\contcaption{ \footnotesize The sources within 3.7$'$ of the galaxy centre. The fluxes are in units of erg\,s$^{-1}$\,cm$^{-2}$ and the luminosities are in units of erg\,s$^{-1}$. The PL and DBB subscripts denote the spectral model conversion factor used to calculate the flux. The "err" designation denotes flux errors. The flag column identifies sources that are interesting and/ or for which caution must be used when analysing the source properties. The key is as follows: bkg = source in an area of high background, c = \emph{wavdetect} has misidentified or not detected the source in one of the observations, cg = the source is in or very close to a chip gap in one of the observations and GC = the source is associated with a globular cluster.} 

\begin{tabular}{|l|l|l|l|l|l|l|l|l|l|l|}
\hline
  \multicolumn{1}{|c|}{RA} &
  \multicolumn{1}{c|}{DEC} &
  \multicolumn{1}{c|}{Flux$_{\rm PL}$} &
  \multicolumn{1}{c|}{Flux$_{\rm PL}$ err} &
  \multicolumn{1}{c|}{Flux$_{\rm PL}$} &
  \multicolumn{1}{c|}{Flux$_{\rm PL}$ err} &
  \multicolumn{1}{c|}{Flux$_{\rm DBB}$} &
  \multicolumn{1}{c|}{Flux$_{\rm DBB}$ err} &
  \multicolumn{1}{c|}{Flux$_{\rm DBB}$} &
  \multicolumn{1}{c|}{Flux$_{\rm DBB}$ err} &
  \multicolumn{1}{c|}{Flag} \\
\midrule
\multicolumn{1}{|c|}{degrees} &
  \multicolumn{1}{c|}{degrees} &
  \multicolumn{1}{c|}{0.5--5\,keV} &
  \multicolumn{1}{c|}{0.5--5\,keV} &
  \multicolumn{1}{c|}{0.5--2\,keV} &
  \multicolumn{1}{c|}{0.5--2\,keV} &
  \multicolumn{1}{c|}{0.5--5\,keV} &
  \multicolumn{1}{c|}{0.5--5\,keV} &
  \multicolumn{1}{c|}{0.5--2\,keV} &
  \multicolumn{1}{c|}{0.5-2\,keV} & 
  \multicolumn{1}{c|}{} \\
\midrule

 187.43102 & 8.03289 & 3.25E-15 & 1.37E-15 & 1.7E-15 & 7.2E-16 & 3.42E-15 & 1.45E-15 & 1.39E-15 & 5.89E-16 & \\
  187.43047 & 8.06057 & 2.41E-15 & 3.56E-16 & 1.26E-15 & 1.87E-16 & 2.54E-15 & 3.77E-16 & 1.03E-15 & 1.53E-16 & \\
  187.43035 & 8.04347 & 3.95E-15 & 1.06E-15 & 2.07E-15 & 5.54E-16 & 4.17E-15 & 1.12E-15 & 1.7E-15 & 4.54E-16 & cg\\
  187.42999 & 7.98177 & 8.77E-16 & 3.02E-16 & 4.6E-16 & 1.58E-16 & 9.25E-16 & 3.18E-16 & 3.76E-16 & 1.3E-16 & \\
  187.42988 & 8.00911 & 1.64E-15 & 3.32E-16 & 8.58E-16 & 1.74E-16 & 1.73E-15 & 3.51E-16 & 7.02E-16 & 1.43E-16 & \\
  187.42951 & 8.01101 & 4.92E-15 & 6.43E-16 & 2.58E-15 & 3.37E-16 & 5.19E-15 & 6.81E-16 & 2.11E-15 & 2.77E-16 & GC\\
  187.42938 & 8.03832 & 5.66E-16 & 2.04E-16 & 2.97E-16 & 1.07E-16 & 5.97E-16 & 2.15E-16 & 2.43E-16 & 8.74E-17 & \\
  187.42923 & 7.98238 & 1.66E-15 & 4.28E-16 & 8.71E-16 & 2.24E-16 & 1.75E-15 & 4.52E-16 & 7.12E-16 & 1.84E-16 & \\
  187.42922 & 7.95775 & 1.85E-15 & 4.78E-16 & 9.73E-16 & 2.51E-16 & 1.96E-15 & 5.04E-16 & 7.95E-16 & 2.05E-16 & \\
  187.42911 & 7.97957 & 2.74E-15 & 4.24E-16 & 1.44E-15 & 2.23E-16 & 2.89E-15 & 4.49E-16 & 1.18E-15 & 1.82E-16 & GC\\
  187.42765 & 8.04481 & 1.11E-15 & 2.89E-16 & 5.81E-16 & 1.52E-16 & 1.17E-15 & 3.06E-16 & 4.75E-16 & 1.24E-16 & cg\\
  187.42639 & 8.00218 & 2.38E-14 & 2.74E-15 & 1.25E-14 & 1.44E-15 & 2.51E-14 & 2.91E-15 & 1.02E-14 & 1.18E-15 & GC\\
  187.42624 & 8.02524 & 8.39E-16 & 2.54E-16 & 4.4E-16 & 1.33E-16 & 8.85E-16 & 2.68E-16 & 3.6E-16 & 1.09E-16 & \\
  187.42603 & 7.95057 & 1.93E-14 & 2.25E-15 & 1.01E-14 & 1.18E-15 & 2.04E-14 & 2.39E-15 & 8.29E-15 & 9.71E-16 & \\
  187.42509 & 8.00396 & 3.43E-15 & 9.75E-16 & 1.8E-15 & 5.12E-16 & 3.62E-15 & 1.03E-15 & 1.47E-15 & 4.19E-16 & \\
  187.42509 & 7.98477 & 2.25E-15 & 5.92E-16 & 1.18E-15 & 3.11E-16 & 2.37E-15 & 6.25E-16 & 9.64E-16 & 2.54E-16 & \\
  187.42493 & 7.98013 & 8.64E-15 & 1.07E-15 & 4.54E-15 & 5.6E-16 & 9.12E-15 & 1.13E-15 & 3.71E-15 & 4.59E-16 & \\
  187.4247 & 8.02168 & 1.21E-15 & 5.16E-16 & 6.37E-16 & 2.71E-16 & 1.28E-15 & 5.45E-16 & 5.21E-16 & 2.22E-16 & \\
  187.42371 & 7.97253 & 2.65E-15 & 5.3E-16 & 1.39E-15 & 2.78E-16 & 2.8E-15 & 5.6E-16 & 1.14E-15 & 2.28E-16 & \\
  187.4236 & 8.00046 & 1.14E-15 & 3.1E-16 & 5.99E-16 & 1.63E-16 & 1.2E-15 & 3.28E-16 & 4.9E-16 & 1.33E-16 & GC\\
  187.4235 & 8.01364 & 1.25E-15 & 2.45E-16 & 6.57E-16 & 1.29E-16 & 1.32E-15 & 2.59E-16 & 5.37E-16 & 1.05E-16 & GC\\
  187.42342 & 8.01731 & 2.82E-15 & 4.04E-16 & 1.48E-15 & 2.12E-16 & 2.97E-15 & 4.28E-16 & 1.21E-15 & 1.74E-16 & \\
  187.42342 & 8.01238 & 1.76E-15 & 3.02E-16 & 9.25E-16 & 1.58E-16 & 1.86E-15 & 3.19E-16 & 7.57E-16 & 1.3E-16 & GC\\
  187.42336 & 8.00398 & 9.62E-15 & 1.18E-15 & 5.05E-15 & 6.17E-16 & 1.01E-14 & 1.25E-15 & 4.13E-15 & 5.07E-16 & GC\\
  187.42334 & 8.01633 & 4.79E-15 & 6.26E-16 & 2.51E-15 & 3.29E-16 & 5.05E-15 & 6.63E-16 & 2.06E-15 & 2.7E-16 & GC\\
  187.42306 & 7.97556 & 3.5E-15 & 5.1E-16 & 1.84E-15 & 2.68E-16 & 3.7E-15 & 5.4E-16 & 1.5E-15 & 2.2E-16 & \\
  187.42266 & 7.98455 & 5.57E-15 & 7.27E-16 & 2.93E-15 & 3.82E-16 & 5.88E-15 & 7.7E-16 & 2.39E-15 & 3.13E-16 & \\
  187.42255 & 7.98166 & 4.62E-15 & 6.39E-16 & 2.43E-15 & 3.35E-16 & 4.88E-15 & 6.77E-16 & 1.98E-15 & 2.75E-16 & \\
  187.42249 & 8.03555 & 1.3E-15 & 2.43E-16 & 6.8E-16 & 1.27E-16 & 1.37E-15 & 2.57E-16 & 5.56E-16 & 1.04E-16 & \\
  187.42242 & 7.95737 & 4.76E-15 & 6.81E-16 & 2.5E-15 & 3.58E-16 & 5.03E-15 & 7.22E-16 & 2.04E-15 & 2.93E-16 & \\
  187.42164 & 7.99188 & 1.54E-15 & 3.51E-16 & 8.09E-16 & 1.84E-16 & 1.63E-15 & 3.71E-16 & 6.61E-16 & 1.51E-16 & \\
  187.42092 & 8.00086 & 3.26E-15 & 4.77E-16 & 1.71E-15 & 2.5E-16 & 3.44E-15 & 5.05E-16 & 1.4E-15 & 2.05E-16 & GC\\
  187.42088 & 7.96233 & 5.11E-14 & 5.78E-15 & 2.68E-14 & 3.03E-15 & 5.39E-14 & 6.13E-15 & 2.19E-14 & 2.49E-15 & GC\\
  187.42039 & 7.99527 & 4.64E-15 & 6.21E-16 & 2.43E-15 & 3.26E-16 & 4.89E-15 & 6.58E-16 & 1.99E-15 & 2.68E-16 & \\
  187.41845 & 8.04004 & 1.04E-15 & 2.24E-16 & 5.45E-16 & 1.18E-16 & 1.1E-15 & 2.37E-16 & 4.45E-16 & 9.64E-17 & \\
  187.4175 & 7.9957 & 2.84E-15 & 4.44E-16 & 1.49E-15 & 2.33E-16 & 3.0E-15 & 4.7E-16 & 1.22E-15 & 1.91E-16 & GC\\
  187.41745 & 7.9748 & 3.9E-15 & 1.23E-15 & 2.04E-15 & 6.46E-16 & 4.11E-15 & 1.3E-15 & 1.67E-15 & 5.29E-16 & GC\\
  187.41707 & 7.99093 & 7.94E-15 & 9.78E-16 & 4.16E-15 & 5.13E-16 & 8.37E-15 & 1.04E-15 & 3.41E-15 & 4.22E-16 & GC\\
  187.41536 & 7.98263 & 1.33E-15 & 3.34E-16 & 6.99E-16 & 1.75E-16 & 1.41E-15 & 3.53E-16 & 5.72E-16 & 1.43E-16 & GC\\
  187.41524 & 7.99126 & 1.67E-15 & 3.86E-16 & 8.76E-16 & 2.03E-16 & 1.76E-15 & 4.08E-16 & 7.16E-16 & 1.66E-16 & GC\\
  187.4145 & 8.02878 & 3.73E-15 & 5.16E-16 & 1.96E-15 & 2.71E-16 & 3.93E-15 & 5.46E-16 & 1.6E-15 & 2.22E-16 & \\
  187.41433 & 8.01041 & 1.96E-15 & 3.36E-16 & 1.03E-15 & 1.77E-16 & 2.07E-15 & 3.56E-16 & 8.43E-16 & 1.45E-16 & \\
  187.41324 & 7.98868 & 5.07E-15 & 6.78E-16 & 2.66E-15 & 3.56E-16 & 5.35E-15 & 7.19E-16 & 2.18E-15 & 2.92E-16 & GC\\
  187.41314 & 7.99475 & 2.74E-15 & 4.25E-16 & 1.44E-15 & 2.23E-16 & 2.9E-15 & 4.5E-16 & 1.18E-15 & 1.83E-16 & GC\\
  187.41275 & 7.95364 & 3.95E-15 & 6.24E-16 & 2.07E-15 & 3.28E-16 & 4.16E-15 & 6.6E-16 & 1.69E-15 & 2.69E-16 & GC\\
  187.41118 & 7.96551 & 5.0E-15 & 7.05E-16 & 2.62E-15 & 3.7E-16 & 5.27E-15 & 7.47E-16 & 2.15E-15 & 3.04E-16 & GC\\
  187.41049 & 8.04092 & 9.77E-16 & 2.48E-16 & 5.13E-16 & 1.3E-16 & 1.03E-15 & 2.62E-16 & 4.19E-16 & 1.06E-16 & \\
  187.41046 & 7.99152 & 1.84E-15 & 6.81E-16 & 9.68E-16 & 3.58E-16 & 1.95E-15 & 7.19E-16 & 7.92E-16 & 2.92E-16 & GC\\
  187.40974 & 8.00833 & 4.49E-15 & 5.95E-16 & 2.36E-15 & 3.13E-16 & 4.74E-15 & 6.31E-16 & 1.93E-15 & 2.57E-16 & \\
  187.40946 & 8.00356 & 1.14E-15 & 2.68E-16 & 5.98E-16 & 1.41E-16 & 1.2E-15 & 2.84E-16 & 4.89E-16 & 1.15E-16 & GC\\
  187.40914 & 8.04296 & 3.24E-15 & 4.74E-16 & 1.7E-15 & 2.49E-16 & 3.42E-15 & 5.02E-16 & 1.39E-15 & 2.04E-16 & \\
  187.40886 & 7.97714 & 4.37E-15 & 7.0E-16 & 2.29E-15 & 3.67E-16 & 4.61E-15 & 7.4E-16 & 1.88E-15 & 3.01E-16 & \\
  187.40746 & 7.97934 & 9.39E-15 & 1.17E-15 & 4.93E-15 & 6.16E-16 & 9.91E-15 & 1.24E-15 & 4.03E-15 & 5.06E-16 & GC\\
  187.40731 & 8.01271 & 4.22E-15 & 5.72E-16 & 2.21E-15 & 3.0E-16 & 4.45E-15 & 6.06E-16 & 1.81E-15 & 2.46E-16 & \\
  187.40654 & 8.02118 & 1.26E-15 & 2.86E-16 & 6.61E-16 & 1.5E-16 & 1.33E-15 & 3.02E-16 & 5.41E-16 & 1.23E-16 & \\
  187.40084 & 7.96944 & 2.1E-15 & 5.37E-16 & 1.1E-15 & 2.82E-16 & 2.21E-15 & 5.67E-16 & 9.0E-16 & 2.31E-16 & \\
  187.40001 & 8.04156 & 1.13E-15 & 3.29E-16 & 5.95E-16 & 1.73E-16 & 1.2E-15 & 3.47E-16 & 4.86E-16 & 1.41E-16 & \\
  187.39989 & 8.01516 & 1.82E-15 & 3.28E-16 & 9.54E-16 & 1.72E-16 & 1.92E-15 & 3.47E-16 & 7.8E-16 & 1.41E-16 & \\
  187.39635 & 7.99997 & 1.59E-15 & 4.39E-16 & 8.33E-16 & 2.3E-16 & 1.67E-15 & 4.63E-16 & 6.81E-16 & 1.88E-16 & \\
  
 \end{tabular}
\end{table*}
\begin{table*}
\footnotesize
\label{source_list}

\contcaption{ \footnotesize The sources within 3.7$'$ of the galaxy centre. The fluxes are in units of erg\,s$^{-1}$\,cm$^{-2}$ and the luminosities are in units of erg\,s$^{-1}$. The PL and DBB subscripts denote the spectral model conversion factor used to calculate the flux. The "err" designation denotes flux errors. The flag column identifies sources that are interesting and/ or for which caution must be used when analysing the source properties. The key is as follows: bkg = source in an area of high background, c = \emph{wavdetect} has misidentified or not detected the source in one of the observations, cg = the source is in or very close to a chip gap in one of the observations and GC = the source is associated with a globular cluster.} 

\begin{tabular}{|l|l|l|l|l|l|l|l|l|l|l|}
\hline
  \multicolumn{1}{|c|}{RA} &
  \multicolumn{1}{c|}{DEC} &
  \multicolumn{1}{c|}{Flux$_{\rm PL}$} &
  \multicolumn{1}{c|}{Flux$_{\rm PL}$ err} &
  \multicolumn{1}{c|}{Flux$_{\rm PL}$} &
  \multicolumn{1}{c|}{Flux$_{\rm PL}$ err} &
  \multicolumn{1}{c|}{Flux$_{\rm DBB}$} &
  \multicolumn{1}{c|}{Flux$_{\rm DBB}$ err} &
  \multicolumn{1}{c|}{Flux$_{\rm DBB}$} &
  \multicolumn{1}{c|}{Flux$_{\rm DBB}$ err} &
  \multicolumn{1}{c|}{Flag} \\
\midrule
\multicolumn{1}{|c|}{degrees} &
  \multicolumn{1}{c|}{degrees} &
  \multicolumn{1}{c|}{0.5--5\,keV} &
  \multicolumn{1}{c|}{0.5--5\,keV} &
  \multicolumn{1}{c|}{0.5--2\,keV} &
  \multicolumn{1}{c|}{0.5--2\,keV} &
  \multicolumn{1}{c|}{0.5--5\,keV} &
  \multicolumn{1}{c|}{0.5--5\,keV} &
  \multicolumn{1}{c|}{0.5--2\,keV} &
  \multicolumn{1}{c|}{0.5-2\,keV} & 
  \multicolumn{1}{c|}{} \\
\midrule

187.39564 & 8.01346 & 4.68E-15 & 6.13E-16 & 2.46E-15 & 3.22E-16 & 4.94E-15 & 6.5E-16 & 2.01E-15 & 2.64E-16 & \\
  187.38648 & 7.98249 & 9.36E-15 & 1.17E-15 & 4.91E-15 & 6.14E-16 & 9.87E-15 & 1.24E-15 & 4.02E-15 & 5.04E-16 & \\
  187.38616 & 7.98217 & 1.31E-14 & 1.75E-15 & 6.88E-15 & 9.21E-16 & 1.38E-14 & 1.86E-15 & 5.62E-15 & 7.56E-16 & \\

 \end{tabular}
\end{table*}

\newpage
%
\makeatletter
\let\@makecaption=\SFB@makefigurecaption
\makeatother
\setlength{\rotFPtop}{0pt plus 1fil}
\setlength{\rotFPbot}{0pt plus 1fil}

\begin{sidewaystable*}
\footnotesize
\label{GC_source_list}

\setlength{\tabcolsep}{5pt}
\footnotesize
\label{GC_source_list}

\caption[The X-ray and optical data for the globular cluster LMXBs]{ \footnotesize The X-ray and optical properties of the GC LMXBs. The X-ray flux and luminosity are for the 0.5--5\,keV range. The X, VCS, RZ and MKZ subscripts denote the RA and DEC for X-ray, RZ01, VCS and MKZ03 observations respectively. \emph{z} and \emph{g} are the photometric values of the sources in the VCS catalogue; \emph{B}, \emph{V} and \emph{R} are the photometric values of the RZ01 catalogue and \emph{V$_{MKZ}$} and \emph{I} are the photometric values for the MKZ03 catalogue.} 
\tiny
\begin{tabular}{|r|r|r|r|l|r|r|r|r|r|r|r|r|r|r|r|r|r|r|r|r|}
\hline
  \multicolumn{1}{|c|}{RA$_{\rm X}$} &
  \multicolumn{1}{c|}{DEC$_{\rm X}$} &
  \multicolumn{1}{c|}{Flux$_{\rm PL}$} &
  \multicolumn{1}{c|}{L$_{\rm X}$} &
  \multicolumn{1}{c|}{Flag} &
  \multicolumn{1}{c|}{RA$_{\rm VCS}$} &
  \multicolumn{1}{c|}{DEC$_{\rm VCS}$} &
  \multicolumn{1}{c|}{$z$} &
  \multicolumn{1}{c|}{$g$} &
  \multicolumn{1}{c|}{$g$-$z$} &
  \multicolumn{1}{c|}{RA$_{\rm RZ}$} &
  \multicolumn{1}{c|}{DEC$_{\rm RZ}$} &
  \multicolumn{1}{c|}{V$_{\rm RZ}$} &
  \multicolumn{1}{c|}{B-V$_{\rm RZ}$} &
  \multicolumn{1}{c|}{V$_{\rm RZ}$-R} &
  \multicolumn{1}{c|}{B-R} &
  \multicolumn{1}{c|}{RA$_{\rm MKZ}$} &
  \multicolumn{1}{c|}{DECmkz03} &
  \multicolumn{1}{c|}{V$_{MKZ}$} &
  \multicolumn{1}{c|}{I} &
  \multicolumn{1}{c|}{V$_{\rm MKZ}$-I} \\
\hline
  degrees & degrees & 0.5-5 keV &  &  & degrees & degrees &  &  &  & degrees & degrees &  &  &  &  & degrees & degrees &  &  & \\
  \hline
  
  187.41745 & 7.9748 & 3.9E-15 & 1.19E38 &  &  &  &  &  &  & 187.41765 & 7.97499 & 22.62 & 0.66 & 0.47 & 1.13 & 187.4175 & 7.975 & 22.77 & 21.8 & 0.97\\
  187.42092 & 8.00086 & 3.26E-15 & 9.98E37 &  & 187.42084 & 8.00083 & 20.9 & 22.29 & 1.4 & 187.42101 & 8.00103 & 21.89 & 0.73 & 0.54 & 1.27 & 187.42083 & 8.00103 & 21.88 & 20.61 & 1.26\\
  187.42336 & 8.00398 & 9.62E-15 & 2.95E38 &  & 187.42325 & 8.00392 & 21.84 & 23.38 & 1.53 & 187.42337 & 8.0041 & 22.88 & 1.21 & 0.55 & 1.76 & 187.42333 & 8.00411 & 22.88 & 21.56 & 1.32\\
  187.42342 & 8.01238 & 1.76E-15 & 5.39E37 &  & 187.42335 & 8.01232 & 20.62 & 22.05 & 1.44 &  &  &  &  &  &  & 187.42333 & 8.01253 & 21.64 & 20.37 & 1.27\\
  187.42334 & 8.01633 & 4.79E-15 & 1.47E38 &  &  &  &  &  &  &  &  &  &  &  &  & 187.42333 & 8.0165 & 23.11 & 22.26 & 0.85\\
  187.42951 & 8.01101 & 4.92E-15 & 1.51E38 &  & 187.42931 & 8.01097 & 20.45 & 21.5 & 1.05 & 187.42945 & 8.01116 & 21.13 & 0.68 & 0.45 & 1.13 & 187.42958 & 8.01117 & 21.11 & 20.1 & 1.01\\
  187.43259 & 7.99911 & 1.52E-15 & 4.65E37 &  & 187.43251 & 7.99904 & 20.42 & 21.61 & 1.2 &  &  &  &  &  &  & 187.4325 & 7.99925 & 21.12 & 20.02 & 1.11\\
  187.43237 & 8.01162 & 2.42E-15 & 7.41E37 &  &  &  &  &  &  &  &  &  &  &  &  & 187.4325 & 8.01153 & 21.2 & 19.9 & 1.29\\
  187.43332 & 7.99764 & 5.77E-15 & 1.77E38 &  & 187.43324 & 7.99757 & 22.26 & 23.61 & 1.35 &  &  &  &  &  &  & 187.43333 & 7.99778 & 23.19 & 21.92 & 1.27\\
  187.43415 & 8.01161 & 3.25E-15 & 9.95E37 &  & 187.43405 & 8.01159 & 19.95 & 21.38 & 1.43 &  &  &  &  &  &  & 187.43417 & 8.01181 & 20.86 & 19.62 & 1.24\\
  187.44078 & 7.96204 & 2.82E-15 & 8.63E37 &  &  &  &  &  &  & 187.44091 & 7.96226 & 21.41 & 0.97 & 0.59 & 1.56 & 187.44083 & 7.96225 & 21.6 & 20.29 & 1.31\\
  187.44202 & 7.98761 & 1.38E-14 & 4.23E38 &  &  &  &  &  &  &  &  &  &  &  &  & 187.44208 & 7.98775 & 23.83 & 22.48 & 1.35\\
  187.44317 & 8.02419 & 3.13E-15 & 9.58E37 &  & 187.44308 & 8.02404 & 23.48 & 24.68 & 1.2 &  &  &  &  &  &  & 187.44333 & 8.02425 & 24.17 & 23.12 & 1.05\\
  187.44595 & 7.97969 & 5.78E-15 & 1.77E38 &  & 187.44588 & 7.97961 & 20.85 & 22.27 & 1.43 &  &  &  &  &  &  & 187.44583 & 7.97983 & 21.79 & 20.55 & 1.24\\
  187.44894 & 7.99055 & 4.14E-15 & 1.27E38 &  & 187.44886 & 7.99043 & 21.74 & 23.23 & 1.49 &  &  &  &  &  &  & 187.44917 & 7.99064 & 22.63 & 21.36 & 1.26\\
  187.44941 & 7.98863 & 2.18E-15 & 6.67E37 &  & 187.4494 & 7.98857 & 20.5 & 21.85 & 1.35 & 187.44954 & 7.98879 & 21.62 & 0.59 & 0.48 & 1.07 & 187.44958 & 7.98878 & 21.33 & 20.13 & 1.2\\
  187.45461 & 8.0076 & 8.41E-16 & 2.57E37 &  & 187.45449 & 8.00753 & 22.08 & 23.42 & 1.34 &  &  &  &  &  &  & 187.45458 & 8.00775 & 22.9 & 21.73 & 1.17\\
  187.45568 & 7.96479 & 5.58E-15 & 1.71E38 &  &  &  &  &  &  & 187.45572 & 7.96497 & 20.82 & 0.71 & 0.52 & 1.23 & 187.45583 & 7.965 & 21.0 & 20.0 & 1.0\\
  187.45878 & 7.99547 & 7.63E-15 & 2.34E38 &  & 187.45878 & 7.99539 & 20.6 & 21.75 & 1.15 & 187.45892 & 7.9956 & 21.3 & 0.76 & 0.51 & 1.27 & 187.45875 & 7.99561 & 21.41 & 20.28 & 1.13\\
  187.46004 & 8.00379 & 9.21E-15 & 2.82E38 &  & 187.45994 & 8.00374 & 20.16 & 21.71 & 1.55 & 187.46008 & 8.00396 & 21.2 & 0.88 & 0.56 & 1.44 & 187.46 & 8.00394 & 21.26 & 19.96 & 1.3\\
  187.462 & 8.00271 & 9.52E-15 & 2.91E38 &  & 187.4619 & 8.00269 & 22.7 & 24.12 & 1.42 &  &  &  &  &  &  & 187.46208 & 8.00289 & 23.61 & 22.45 & 1.16\\
  187.46394 & 7.98809 & 2.08E-15 & 6.37E37 &  &  &  &  &  &  & 187.46397 & 7.98823 & 21.99 & 0.85 & 0.54 & 1.39 & 187.46417 & 7.98825 & 21.75 & 20.57 & 1.18\\
  187.46556 & 7.99996 & 4.17E-15 & 1.28E38 &  & 187.46546 & 7.99993 & 19.36 & 20.8 & 1.44 &  &  &  &  &  &  & 187.46542 & 8.00014 & 20.39 & 19.14 & 1.25\\
  187.46883 & 8.00051 & 3.14E-15 & 9.61E37 &  & 187.46873 & 8.00045 & 22.63 & 24.2 & 1.57 &  &  &  &  &  &  & 187.46875 & 8.00067 & 23.56 & 22.35 & 1.21\\
  187.40746 & 7.97934 & 9.39E-15 & 2.87E38 &  &  &  &  &  &  & 187.40752 & 7.97947 & 22.03 & 0.75 & 0.49 & 1.24 &  &  &  &  & \\
  187.41118 & 7.96551 & 5E-15 & 1.53E38 &  &  &  &  &  &  & 187.41133 & 7.96565 & 22.94 & 0.84 & 0.42 & 1.26 &  &  &  &  & \\
  187.41275 & 7.95364 & 3.95E-15 & 1.21E38 &  &  &  &  &  &  & 187.41284 & 7.95371 & 21.43 & 0.86 & 0.55 & 1.41 &  &  &  &  & \\
  187.41324 & 7.98868 & 5.07E-15 & 1.55E38 &  & 187.41309 & 7.98874 & 19.91 & 21.29 & 1.38 & 187.41327 & 7.98894 & 20.88 & 0.79 & 0.57 & 1.36 &  &  &  &  & \\
  187.4175 & 7.9957 & 2.84E-15 & 8.7E37 &  & 187.41742 & 7.99564 & 19.92 & 21.23 & 1.31 & 187.4176 & 7.99583 & 20.82 & 0.8 & 0.56 & 1.36 &  &  &  &  & \\
  187.42088 & 7.96233 & 5.11E-14 & 1.56E39 &  &  &  &  &  &  & 187.42096 & 7.96246 & 21.55 & 0.85 & 0.59 & 1.44 &  &  &  &  & \\
  187.4235 & 8.01364 & 1.25E-15 & 3.83E37 &  & 187.4234 & 8.01364 & 21.3 & 22.33 & 1.04 & 187.42356 & 8.01384 & 22.06 & 0.46 & 0.34 & 0.8 &  &  &  &  & \\
  187.4236 & 8.00046 & 1.14E-15 & 3.49E37 &  & 187.42352 & 8.00044 & 19.84 & 21.51 & 1.67 & 187.42369 & 8.00065 & 21.01 & 0.95 & 0.61 & 1.56 &  &  &  &  & \\
  187.42639 & 8.00218 & 2.38E-14 & 7.29E38 &  & 187.42631 & 8.00214 & 19.57 & 21.0 & 1.43 & 187.42647 & 8.00233 & 20.55 & 0.82 & 0.55 & 1.37 &  &  &  &  & \\
  187.42911 & 7.97957 & 2.74E-15 & 8.39E37 &  & 187.42901 & 7.97948 & 20.36 & 21.72 & 1.36 & 187.42918 & 7.97969 & 21.23 & 0.85 & 0.44 & 1.29 &  &  &  &  & \\
  187.43269 & 7.96569 & 3.06E-15 & 9.37E37 &  &  &  &  &  &  & 187.43294 & 7.96593 & 20.59 & 0.87 & 0.54 & 1.41 &  &  &  &  & \\
  187.44648 & 7.97605 & 1.78E-15 & 5.45E37 &  & 187.44634 & 7.97604 & 20.59 & 22.25 & 1.66 & 187.4465 & 7.97625 & 21.69 & 0.83 & 0.56 & 1.39 &  &  &  &  & \\
  187.44716 & 8.01575 & 1.04E-15 & 3.18E37 &  & 187.44701 & 8.01579 & 20.41 & 21.62 & 1.2 & 187.44718 & 8.01601 & 21.41 & 0.66 & 0.55 & 1.21 &  &  &  &  & \\
  187.44753 & 7.98658 & 1.1E-15 & 3.37E37 &  & 187.44736 & 7.98646 & 21.26 & 22.64 & 1.39 & 187.44752 & 7.98668 & 22.47 & 0.63 & 0.36 & 0.99 &  &  &  &  & \\
  187.45009 & 7.97628 & 6.51E-16 & 1.99E37 &  & 187.44992 & 7.97634 & 20.41 & 21.7 & 1.29 & 187.45009 & 7.97657 & 21.371 & 0.71 & 0.43 & 1.14 &  &  &  &  & \\
  187.45507 & 8.0155 & 7.23E-16 & 2.21E37 &  & 187.45493 & 8.01541 & 20.37 & 21.87 & 1.49 & 187.4551 & 8.01561 & 21.54 & 0.73 & 0.57 & 1.3 &  &  &  &  & \\
  187.4555 & 8.03852 & 1.23E-15 & 3.77E37 &  &  &  &  &  &  & 187.45559 & 8.03869 & 20.59 & 0.86 & 0.59 & 1.45 &  &  &  &  & \\
  187.45636 & 7.99033 & 1.62E-15 & 4.96E37 & bkg & 187.45633 & 7.99025 & 20.55 & 22.05 & 1.5 & 187.45652 & 7.99047 & 21.81 & 0.88 & 0.65 & 1.53 &  &  &  &  & \\
  187.46276 & 7.99645 & 3.37E-15 & 1.03E38 &  & 187.46286 & 7.99632 & 19.81 & 21.28 & 1.48 & 187.46303 & 7.99654 & 20.98 & 0.88 & 0.57 & 1.45 &  &  &  &  & \\
  187.46914 & 8.00449 & 1.65E-15 & 5.05E37 &  & 187.46901 & 8.00449 & 20.67 & 22.12 & 1.45 & 187.46917 & 8.00467 & 21.77 & 0.8 & 0.55 & 1.35 &  &  &  &  & \\
  187.46969 & 7.95261 & 1.78E-15 & 5.45E37 &  &  &  &  &  &  & 187.46981 & 7.95284 & 21.82 & 0.64 & 0.42 & 1.06 &  &  &  &  & \\
  187.47006 & 7.97202 & 2.93E-15 & 8.97E37 &  &  &  &  &  &  & 187.4701 & 7.97213 & 20.97 & 0.83 & 0.59 & 1.42 &  &  &  &  & \\
  187.47223 & 7.99779 & 3.61E-15 & 1.11E38 &  &  &  &  &  &  & 187.47231 & 7.99785 & 21.98 & 0.53 & 0.3 & 0.83 &  &  &  &  & \\
  187.47234 & 8.00724 & 3.51E-15 & 1.07E38 &  & 187.47224 & 8.0072 & 19.08 & 20.54 & 1.47 & 187.47238 & 8.00741 & 20.03 & 0.86 & 0.56 & 1.42 &  &  &  &  & \\
  187.47229 & 8.0526 & 3.17E-15 & 9.71E37 &  &  &  &  &  &  & 187.47238 & 8.05279 & 22.01 & 0.76 & 0.58 & 1.34 &  &  &  &  & \\
  187.47602 & 8.00271 & 4.86E-15 & 1.49E38 &  &  &  &  &  &  & 187.47611 & 8.00288 & 21.34 & 0.55 & 0.46 & 1.01 &  &  &  &  & \\
  187.48287 & 8.00096 & 1.57E-15 & 4.81E37 &  &  &  &  &  &  & 187.48287 & 8.00112 & 21.89 & 0.92 & 0.57 & 1.49 &  &  &  &  & \\
  187.48301 & 7.95625 & 2.27E-15 & 6.95E37 &  &  &  &  &  &  & 187.48301 & 7.95649 & 20.74 & 0.93 & 0.58 & 1.51 &  &  &  &  & \\
  187.48483 & 8.00848 & 4.12E-15 & 1.26E38 &  &  &  &  &  &  & 187.48489 & 8.00865 & 21.86 & 0.95 & 0.6 & 1.55 &  &  &  &  & \\
  187.49021 & 7.9924 & 1.71E-15 & 5.24E37 &  &  &  &  &  &  & 187.49016 & 7.99268 & 21.44 & 0.61 & 0.51 & 1.12 &  &  &  &  & \\
  187.40946 & 8.00356 & 1.14E-15 & 3.49E37 &  & 187.40924 & 8.00342 & 21.22 & 22.62 & 1.4 &  &  &  &  &  &  &  &  &  &  & \\
  187.41046 & 7.99152 & 1.84E-15 & 5.63E37 &  & 187.41033 & 7.99147 & 21.39 & 22.83 & 1.44 &  &  &  &  &  &  &  &  &  &  & \\
  187.41314 & 7.99475 & 2.74E-15 & 8.39E37 &  & 187.41307 & 7.9947 & 21.61 & 23.05 & 1.44 &  &  &  &  &  &  &  &  &  &  & \\
  187.41524 & 7.99126 & 1.67E-15 & 5.11E37 &  & 187.4151 & 7.99117 & 21.43 & 22.61 & 1.18 &  &  &  &  &  &  &  &  &  &  & \\
  187.41536 & 7.98263 & 1.33E-15 & 4.07E37 &  & 187.41523 & 7.98275 & 23.42 & 24.74 & 1.32 &  &  &  &  &  &  &  &  &  &  & \\
  187.41707 & 7.99093 & 7.94E-15 & 2.43E38 &  & 187.41699 & 7.99086 & 20.44 & 22.06 & 1.62 &  &  &  &  &  &  &  &  &  &  & \\
  187.43343 & 8.02091 & 1.92E-15 & 5.88E37 &  & 187.43337 & 8.0209 & 21.12 & 22.78 & 1.66 &  &  &  &  &  &  &  &  &  &  & \\
  187.43375 & 7.97804 & 6.34E-15 & 1.94E38 &  & 187.43367 & 7.97794 & 22.84 & 24.45 & 1.62 &  &  &  &  &  &  &  &  &  &  & \\
  187.43483 & 7.99499 & 9.4E-16 & 2.88E37 &  & 187.43471 & 7.99488 & 21.88 & 23.53 & 1.65 &  &  &  &  &  &  &  &  &  &  & \\
  187.43763 & 7.99741 & 3.48E-15 & 1.07E38 &  & 187.43756 & 7.99727 & 20.82 & 22.37 & 1.55 &  &  &  &  &  &  &  &  &  &  & \\
  187.43899 & 7.99449 & 2.04E-15 & 6.25E37 &  & 187.43886 & 7.99442 & 21.95 & 23.6 & 1.65 &  &  &  &  &  &  &  &  &  &  & \\
  187.4391 & 7.99342 & 1.71E-15 & 5.24E37 &  & 187.43902 & 7.9934 & 20.43 & 21.88 & 1.45 &  &  &  &  &  &  &  &  &  &  & \\
  187.44004 & 7.99341 & 7.4E-16 & 2.27E37 &  & 187.44 & 7.99333 & 22.62 & 24.06 & 1.44 &  &  &  &  &  &  &  &  &  &  & \\
  187.44023 & 7.99462 & 1.04E-15 & 3.18E37 &  & 187.44014 & 7.99453 & 21.83 & 23.06 & 1.23 &  &  &  &  &  &  &  &  &  &  & \\
  187.44131 & 8.02414 & 5.76E-15 & 1.76E38 &  & 187.4412 & 8.02413 & 23.73 & 25.09 & 1.37 &  &  &  &  &  &  &  &  &  &  & \\
  187.44205 & 8.02309 & 1.15E-15 & 3.52E37 &  & 187.44214 & 8.02292 & 22.12 & 23.53 & 1.41 &  &  &  &  &  &  &  &  &  &  & \\
  187.44231 & 7.98961 & 3.31E-15 & 1.01E38 &  & 187.44223 & 7.98957 & 23.22 & 23.76 & 0.54 &  &  &  &  &  &  &  &  &  &  & \\
  187.44982 & 8.00233 & 8.39E-15 & 2.57E38 & bkg & 187.44972 & 8.00222 & 19.41 & 20.82 & 1.41 &  &  &  &  &  &  &  &  &  &  & \\
  187.45034 & 7.99476 & 1.44E-15 & 4.41E37 &  & 187.45024 & 7.99478 & 19.37 & 20.62 & 1.25 &  &  &  &  &  &  &  &  &  &  & \\
  187.45379 & 7.97948 & 1.73E-15 & 5.3E37 &  & 187.45386 & 7.97938 & 21.2 & 22.7 & 1.5 &  &  &  &  &  &  &  &  &  &  & \\
  187.45612 & 7.99293 & 1.86E-15 & 5.69E37 &  & 187.45606 & 7.99289 & 20.88 & 22.31 & 1.42 &  &  &  &  &  &  &  &  &  &  & \\
  187.46132 & 8.01972 & 1.18E-15 & 3.61E37 &  & 187.46133 & 8.01967 & 22.69 & 24.33 & 1.65 &  &  &  &  &  &  &  &  &  &  & \\
  187.46225 & 7.98736 & 1.49E-15 & 4.56E37 &  & 187.46202 & 7.98729 & 21.62 & 23.22 & 1.6 &  &  &  &  &  &  &  &  &  &  & \\
  187.46543 & 7.99196 & 1.39E-15 & 4.26E37 &  & 187.46522 & 7.9919 & 22.49 & 23.9 & 1.41 &  &  &  &  &  &  &  &  &  &  & \\
  187.46582 & 7.99676 & 3.14E-15 & 9.61E37 &  & 187.46584 & 7.99667 & 22.4 & 23.93 & 1.53 &  &  &  &  &  &  &  &  &  &  & \\
  187.46921 & 8.01933 & 1.96E-15 & 6.0E37 &  & 187.46911 & 8.01926 & 22.85 & 24.32 & 1.47 &  &  &  &  &  &  &  &  &  &  & \\
\hline
\end{tabular}
\end{sidewaystable*}

 \end{document}